\documentclass[aps,pre,superscriptaddress,twocolumn,showpacs,floatfix]{revtex4}

\usepackage{amsfonts}
\usepackage{epsfig}
\usepackage{graphicx}
\usepackage{subfigure}

\newcommand{\snm}[1]{s_{#1}^{-}}
\newcommand{\snp}[1]{s_{#1}^{+}}
\newcommand{\snz}[1]{s_{#1}^{0}}
\newcommand{\uo}[1]{\underline{\omega}_{#1}}

\begin{document}
%----------------------------------------------------------------------------------------------
\title{Fractality of the non-equilibrium stationary states of open\\
  volume-preserving systems: II. Galton boards}
%----------------------------------------------------------------------------------------------
\author{Felipe Barra}
\affiliation{Departamento de F\'{\i}sica, Facultad de Ciencias
F\'{\i}sicas y Matem\'aticas, Universidad de Chile, Casilla 487-3,
Santiago Chile}
\author{Pierre Gaspard}
\affiliation{Center for Nonlinear Phenomena and Complex Systems,
  Universit\'e Libre  de Bruxelles, C.~P.~231, Campus Plaine, B-1050
  Brussels, Belgium}
\author{Thomas Gilbert}
\affiliation{Center for Nonlinear Phenomena and Complex Systems,
  Universit\'e Libre  de Bruxelles, C.~P.~231, Campus Plaine, B-1050
  Brussels, Belgium}
\date{\today}
%----------------------------------------------------------------------------------------------
\begin{abstract}
Galton boards are models of deterministic diffusion in a uniform
external field, akin to driven periodic Lorentz gases, here
considered in the absence of dissipation mechanism. Assuming a
cylindrical geometry with axis along the direction of the external field,
the two-dimensional board becomes a model for one-dimensional mass
transport along the direction of the external field. This is a purely
diffusive process which admits fractal non-equilibrium  stationary states
under flux boundary conditions.  Analytical results are obtained for the
statistics of multi-baker maps modeling such a non-uniform diffusion
process. A correspondence is established between the local phase-space
statistics and their macroscopic counter-parts. The fractality of the
invariant state is shown to be responsible for the positiveness of the 
entropy production rate.
\end{abstract}

\pacs{05.45.-a,05.70.Ln,05.60.-k}
%----------------------------------------------------------------------------------------------
\maketitle
%----------------------------------------------------------------------------------------------
\section{\label{sec.int}Introduction}

Studying the statistical properties of simple mechanical models with
strongly chaotic dynamics helps understanding the connection between
deterministic motion at the microscopic scale and transport processes which
occur at the macroscopic scales. This is of particular importance with
regards to the irreversibility of thermodynamics and specifically the
dynamical origins of the positiveness of entropy production. 

Such a mechanical device was originally introduced by Sir Francis Galton in
the form of an apparatus which provides a mechanical illustration of the
Gaussian spreading of independent random events \cite{Galton1889}. The
Galton board, also known as quincunx or bean machine \cite{mathworld},
consists of an upright board with a periodic array of pegs upon which a
charge of small shots is released. The particles are let to collide on
the way downward, thus displaying a seemingly erratic motion through the
successive rows of pegs, until they reach the bottom of the board, where
they are stopped.  

Provided the actual dynamics are sufficiently chaotic and dissipative, one
can idealize 
individual paths as Bernoulli trials, whereby every collision event results
into the pellets hopping down to the right or left of the pegs with equal 
probabilities. The number of steps in the trials is then specified by the
number of the rows of pegs in the board. Under such conditions, the heaps
of shots that form at the bottom of the board are expected to be
distributed according to a binomial distribution and thus approximate a
normal distribution. 

Though Galton's board was intended precisely as a mechanical illustration
of this idealized model, the dynamics of the board are necessarily more
intricate, in particular with regards to inelasticity of the collisions
between pegs and pellets and the friction exerted by the board's surface on
the pellets. However if the collisions between the pellets and
pegs were perfectly elastic and the board frictionless, the energy of every
individual pellet would be conserved along its path. As a consequence, the
kinetic energy would increase linearly with the distance separating 
the pellet position from the top of the board, where one can assume
it was released with a specified velocity, which, for the sake
of specializing the motion to a fixed energy shell, we assume to be equal
in magnitudes for all pellets. Such a conservative Galton board is
also referred to as idealized.

The remarkable property of conservative Galton boards is that a
pellet's motion is recurrent, which is contrary to what had until recently
seemed to be a widespread consensus. In other words, however far a pellet
goes in the direction of  the external field, and consequently however large
its kinetic energy becomes, it will come back to the top of the board with
probability one.  This property was proved by Chernov and Dolgopyat
\cite{CD07a,CD07b}, who also showed, in accordance to previous 
heuristic arguments and numerical studies, that the presence of the
external field affects the scaling law of positions and velocities so that
a pellet's speed scales according to $v(t)\sim t^{1/3}$ and its coordinate
$x(t)\sim t^{2/3}$. They further found exact limit distributions for the
rescaled velocity $t^{-1/3}v(t)$ and position $t^{-2/3}x(t)$. 

Galton boards and related models have attracted much attention in the
statistical physics community. In particular, Lorentz gases, which describe
the motion of independent classical point particles in an 
array of fixed scattering disks, have been the subject of intensive 
investigations as models of diffusive transport of light tracer particles
among heavier ones \cite{Lor1905,vB82,Det00,CC70,HKM74}. Lorentz
gases have enjoyed a privileged status in the development of
non-equilibrium statistical mechanics, which stems from the simplicity of
its dynamics. By neglecting the recoil of heavy particles upon collision
with the light tracer particles, one obtains a low-dimensional model that
is amenable to a proper thermodynamical treatment while it retains
important characteristics of genuine many-particle systems. This
model has been studied with mathematical rigor and, in particular, the
existence of a well-defined diffusion coefficient has been proved
rigorously under certain conditions \cite{BS80}. Furthermore, in the last
decades, and in the context of molecular dynamics simulations of
non-equilibrium systems \cite{Hoo86,   EM90}, several versions of the
Lorentz gas model have been considered, including the  Gaussian
thermostated Lorentz gas in the presence of a uniform external field
\cite{MH87}, for which the Einstein relation between the coefficients of
electrical conductivity and diffusion has been proved \cite{CELS93}.

The reason for the initial success of the Lorentz gas was its use by
Lorentz \cite{Lor1905}, elaborating on Drude's theory of electrical and
thermal conduction \cite{Dru1900,Mer76}, for the sake of deriving the
Wiedemann-Franz law, which predicts the temperature  dependence of the
ratio between heat and electrical conductivities in metals. 
In this framework, the computation of the electrical conductivity
assumes that the external field is weak enough that the tracer particle
velocity magnitude is constant. Thus the diffusion coefficient is
homogeneous and essentially given by the product of the particle's mean free
path and (thermal) velocity.

In a conservative diffusive system acted upon by an external field, the
situation is different in that the external field causes the acceleration
of particles and induces a velocity-dependent diffusion
coefficient. Nevertheless such a system bears strong analogies with the
field free diffusive case. 

It is our purpose to investigate this analogy by comparing the statistical
properties of Galton boards to that of periodic Lorentz gases. The latter
were studied in a first paper \cite{BGGi}, where we discussed the
fractality of the non-equilibrium stationary states of open Lorentz gases
under flux boundary conditions, \emph{i.~e.} a slab of finite
extension with its two boundaries in contact with particle reservoirs
with differing injection rates. Under such boundary conditions, the
Lorentz gas sustains a steady current of mass which induces a constant rate
of entropy production. 

In \cite{BGGi}, we established the 
connection between this production of entropy and the fractality of the
stationary states of open Lorentz gases. In
this follow-up paper, we extend these results to Galton boards and related
models. In particular, we develop a discrete random walk model that mimics
the collision dynamics of Galton boards and associate to it a multi-baker
map with energy, similar to models introduced in \cite{TG99, TG00}. The
specificity of our model is that the transition rates vary
with the sites' indices, reflecting the property of the conservative Galton
board that deflection of tracers by the external field is more likely
to occur when their kinetic energies are small in comparison to their
potential energies. We derive the analytic expression of the
non-equilibrium stationary states of this multi-baker map and show that its
cumulative measures are characterized by nowhere differentiable continuous
functions similar to the Takagi function of the non-equilibrium stationary
state of the usual multi-baker map \cite{TG95}. This allows us 
to compute the entropy associated to such non-equilibrium stationary states
and thus obtain an analytic derivation of the rate of entropy production,
which, within our formalism, finds its origin in the fractality of the 
non-equilibrium stationary state, in agreement with the results
presented in \cite{BGGi} for the field free case.

The paper is organized as follows. Galton boards are presented in
Sec.~\ref{sec.gbd}. The connection to the phenomenology of diffusion in an
external field, described in Sec.~\ref{subsec.phy}, is established for
both closed and open systems, whose statistical properties are considered
in Sec.~\ref{subsec.dzd}. In Sec.~\ref{subsec.isl} we discuss the
occurrence of elliptic islands in the Galton board's dynamics, \emph{i.e.}
the stabilization of periodic orbits, and provide conditions under which we
can assume the system to be fully hyperbolic. This regime is studied
numerically, first under equilibrium setting in Sec.~\ref{subsec.egb}, and
then under non-equilibrium setting in Sec.~\ref{subsec.ngb}, where we
demonstrate the fractality of the invariant measure. In
Sec.~\ref{sec.fmb}, we introduce the forced multi-baker map, which mimics
the collision dynamics of the Galton board and analyze its statistics in
Sec.~\ref{subsec.ste}. The entropy production rate of the
non-equilibrium stationary state is computed in Sec.~\ref{subsec.ep}. We
end with conclusions in Sec.~\ref{sec.con}.

\section{\label{sec.gbd}Galton board}

The Galton board is similar to a periodic Lorentz gas in a uniform field.
We consider a two-dimensional cylinder of length $L = Nl$ and height
$\sqrt{3}l$, with disks $\mathbb{D}_n$, $0\leq n\leq 2 N$, of radii
$\sigma$,  $\sqrt{3}/4<\sigma/l<1/2$, placed on a hexagonal lattice
structure. The centers of the disks take positions 
\begin{equation}
  (x_n,y_n) = 
  \left\{
    \begin{array}{l@{\quad}l}
      (nl/2, 0)\,,&n\,\mathrm{odd},\\
      (n l/2, \pm \sqrt{3}l/2)\,,& n\,\mathrm{even},
    \end{array}
  \right.
\end{equation}
where we identify the disks $y = \pm \sqrt{3}l/2$. 
The cylindrical region around disk $\mathbb{D}_n$ is defined as
\begin{equation}
  \mathbb{I}_n = \big\{(x,y)\ |\ (n - 1/2)l/2 \leq x \leq (n + 1/2)l/2
  \big\}.
\end{equation}
Thus the interior of the cylinder, where particles propagate freely is made
up of the union $\cup_{n = -N}^N \mathbb{I}_n \setminus \mathbb{D}_n$.

The associated phase space, defined on a constant energy shell, is
$\mathbb{C} = \cup_{n=-N}^N \mathbb{C}_n$, where $\mathbb{C}_n =
\mathbb{S}^1 \otimes [\mathbb{I}_n \setminus \mathbb{D}_n]$ and the unit
circle $\mathbb{S}^1$ represents all possible velocity
directions. Particles are reflected with  elastic collision rules on the
border $\partial \mathbb{C}$, except at the external borders, corresponding
to $x = 0, L$, where they get absorbed. 
Points in phase-space are denoted by $\Gamma = (x,y,v_x,v_y)$, and
trajectories by  $\Phi^t\Gamma$, with $\Phi^t$ the flow associated to the
dynamics of the Galton board. 

The collision map takes the point $\Gamma = (x, y,
v_x, v_y) \in \partial \mathbb{C}$ to $\Phi^\tau \Gamma = (x',y',v_x',v_y')
\in \partial \mathbb{C}$, where $\tau$ 
is the time that separates the two successive collisions with the border
of the Lorentz channel $\partial \mathbb{C}$, and $(v_x',v_y')$ is obtained
from $(v_x, v_y)$ first by propagation under the uniform accelaration until
the instant of collision, and then applying the usual rules of specular
collisions. Given 
that the energy $E$ is fixed, the collision map operates on a
two-dimensional surface, which, when the collision takes place on
disk $n$, is conveniently parameterized by the Birkhoff
coordinates $(\phi_n, \xi_n)$, where $\phi_n$ specifies a
generalized angle variable along the border of disk $n$, to be determined
in Sec. \ref{subsec.egb}, and $\xi_n$ is the sinus of the angle that the
particle velocity makes with respect to the outgoing normal to the disk
after the collision. 

The external field is uniform and directed
along the positive $x$ direction, so that particles accelerate as they move
along the axis of the channel, in the direction of the external
field. There is no dissipative mechanism and energy is conserved along
the Galton board trajectories. 

In this system, as opposed to typical billiards, the energy, denoted $E$,
can be both kinetic and potential. As the particle moves along the
direction of the channel axis, it looses potential energy and gains kinetic
energy, according to the energy conservation $E = (v_x^2+v_y^2)/2 -
\epsilon x$, where $\epsilon$ denotes the amplitude of the external
field. Conversely, the particle looses kinetic energy and gains potential
energy as it moves in the direction opposite to the external field. 

Assuming $E\ge0$, the boundaries of the system are placed
at  $x=0$ and $x=L$, reflecting the impossibility for a trajectory to gain
potential energy beyond the zero kinetic energy level. When $E=0$,
trajectories turn around at $x\ge0$ when the $x$ component of the velocity
annihilates, whereas when $E>0$, depending on the choice of boundary
conditions, particles can be either reflected or absorbed when they reach
$x=0$. 

The trajectory between two successive elastic collisions with the disks is
now parabolic, according to $x(t) = x(0) + v_x(0)t + \epsilon 
t^2/2$, whereas the vertical motion is uniform $y(t) = y(0) + v_y(0)t$. 
The amplitude of the external field $\epsilon$ can be set to
unity by an appropriate rescaling of the momenta and time variable: $v\to
v/\sqrt{\epsilon}$ and $t\to t\sqrt{\epsilon}$. Correspondingly, the energy
has the units of length.

We can thus write the velocity amplitude as a function of the $x$
coordinate, 
\begin{equation}
  v(x) = \sqrt{2(E + x)}.
  \label{gbdpx}
\end{equation}
In particular, the velocity amplitude at $x=0$ is $v(0) = \sqrt{2E}$.
We will assume that the energy takes half integer values of the cell widths
$l$, so that the kinetic energy takes half integer values at the horizontal
positions of the disks along the channel, \emph{i.e} at
$x$'s which are half integer multiples of $l$. 

\begin{figure*}[bth]
  \centering
  \includegraphics[width=\textwidth]{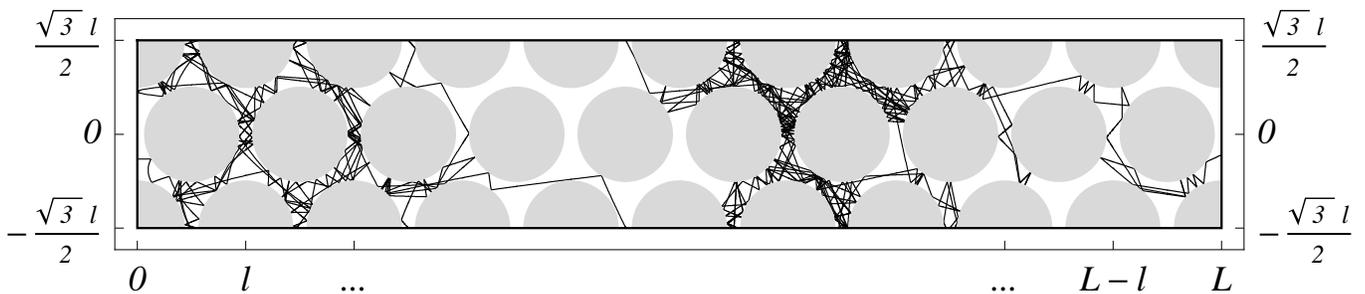}
  \caption{Cylindrical Galton board with a non-vanishing external field and
    the energy $E=0$. A trajectory is
    released at zero velocity at $x=0$, and falls along the external field
    until it collides a first time with a disk. It then wanders around,
    coming back close to $x=0$ once, after which it moves further along the
    channel until it reaches the border at $x=L$. To compute the successive
    collision events, the numerical integration scheme uses an exact
    quartic equation solver based on the Galois formula \cite{WQu}.} 
  \label{fig.gbd}
\end{figure*}

The system is shown in Fig.~\ref{fig.gbd} with absorbing boundary
conditions at $x=0$ and $L$. Note that trajectories are seen to bend
along the field only so long as the velocity is small enough that the
action of the field is noticeable. Otherwise the trajectory 
looks much like that of the Lorentz channel in the absence of external
field. The time scales are however different.

\subsection{\label{subsec.phy}Phenomenology}

One often reads in the literature that Galton boards, or equivalently
periodic Lorentz gases in a uniform external field, do not have a
stationary state. This is however a confusing statement since the existence
of the stationary state has nothing to do with the presence of the external
field. Rather, it is a matter of boundary conditions. 

Just as with the usual Lorentz gas, when an external forcing is turned on,
a stationary state is reached so long as one specifies the boundary
conditions. The reason 
for much of the confusion associated to this problem is, according to our
understanding, that one cannot consider periodic boundary conditions along
the direction of the field since they would violate the conservation of
energy. One can however consider both reflecting and absorbing boundary
conditions for the extended system. The nature of the stationary state,
whether equilibrium or non-equilibrium, depends on the choice of boundary
conditions.

A phenomenological diffusion equation can be obtained for the motion along
the axis of the cylindrical channel, which corresponds to the direction of
the external field. 

In the presence of an external field, the diffusion process is a priori
biased, so that the Fokker-Planck equation of diffusion reads
\begin{equation}
  \partial_t \mathcal{P}(X,t) = \partial_X [\mathcal{D}(X) 
  \partial_X \mathcal{P}(X,t) + \mathcal{M}(X) \mathcal{P}(X,t)]\,.
  \label{bFPeq}
\end{equation}
Here $X$ denotes a macroscopic position, associated to the projection along
the axis direction of a given phase-space region $\mathbb{I}_n$ of the
Galton board, taken in the continuum limit.

According to Einstein's argument, the diffusion coefficient
$\mathcal{D}(X)$ is connected to the mobility coefficient $\mathcal{M}(X)$
by the condition that Eq. (\ref{bFPeq}) admits the equilibrium state
$\mathcal{P}_\mathrm{eq}(X)$ as a solution which annihilates the mean
current:
\begin{equation}
  \mathcal{D}(X) \partial_X \mathcal{P}_\mathrm{eq}(X) 
  + \mathcal{M}(X) \mathcal{P}_\mathrm{eq}(X) = 0\,.
  \label{zeroeqcurrent}
\end{equation}
At the microscopic level, letting $\Gamma$ denote a phase point in $2d$
dimensions with velocity amplitude $v$ and position $x$ with respect to the
direction of the external field, the equilibrium state is the
microcanonical state
\emph{i.~e.} 
\begin{equation}
  \rho_\mathrm{eq}(\Gamma) \propto %=  \mathcal{N}
  \delta \left(E - \frac{v^2}{2} + x \right)\,,
  \label{microeq}
\end{equation}
Integrating this equilibrium phase-space density over cells $\mathbb{C}_n$
and taking the continuum limit $l\to0$ and $n\to\infty$ with the
macroscopic position variable $X = n l/2$ fixed, we obtain the macroscopic
equilibrium density 
$\mathcal{P}_\mathrm{eq}(X)$, 
\begin{eqnarray}
  \mathcal{P}_\mathrm{eq}(X) dX&=& 
  \lim_{\stackrel{l\to0}{n\to\infty}}
  \int_{\mathbb{C}_n}d\Gamma
  \rho_\mathrm{eq}(\Gamma)\,,\nonumber\\
  &\propto&  
  \lim_{\stackrel{l\to0}{n\to\infty}} \int_{\mathbb{C}_n}d\Gamma
  \delta \left(E - \frac{v^2}{2} + x \right)\,,
  \nonumber\\
  &\propto&
  %\int_{(n - 1/2)l/2}^{(n + 1/2)l/2} dx\,
  \lim_{l\to0} l \int dv\,v^{d-1} \delta \left(E - \frac{v^2}{2} + X \right)\,.
  \label{ccP}
\end{eqnarray}
Identifying the length increments $dX = l$, and carrying out the velocity
integration, we arrive to the expression of the equilibrium density
\begin{equation}
  \mathcal{P}_\mathrm{eq}(X) = \mathcal{N} [2(E + X)]^{\frac{d - 2}{2}}\,,
  \label{macroPeq}
\end{equation}
where $\mathcal{N}$ is a normalization factor. 
Inserting this expression into Eq.~(\ref{zeroeqcurrent}), we obtain the
relation between the mobility and diffusion coefficients,
\begin{equation}
  \mathcal{M}(X) = - \frac{d - 2}{2} \frac{\mathcal{D}(X)}{E + X}\,.
  \label{mobilitydiffusion}
\end{equation}

The diffusion coefficient, on the other hand, is proportional to the
magnitude of the position-dependent velocity, $V(X) = \sqrt{2(E + X)}$. This
is a transposition of the corresponding result for the usual field free
periodic Lorentz gas, where the tracer's velocity has constant magnitude.
In the Galton board, given an energy $E$ identical for all the tracer
particles, the velocities $V(0) = \sqrt{2E}$ at $X=0$ are identical for all
particles, growing with $X>0$, due to the uniform force of unit amplitude
acting along that direction. We can therefore write
\begin{equation}
  \mathcal{D}(X) = \mathcal{D}_0 \sqrt{1+\frac{X}{E}}
  \label{fDiff}
\end{equation}
Notice the normalization so chosen that the
diffusion coefficient at $X = 0$ reduces to $\mathcal{D}_0$.
Equation (\ref{fDiff}) can be thought of as a transposition of the argument
by Machta and Zwanzig \cite{MZ83} who provided an analytical expression of
the diffusion coefficient for the periodic Lorentz gas, based upon a random
walk approximation. This approximation indeed carries over to the Galton
board. Provided energy is conserved, the velocity of a tracer 
particle increases as it moves along the direction of the external field.
Thus, provided the periodic cells have sizes small enough that
velocities remain approximately constant within each cell, the
Machta-Zwanzig argument tells us that the diffusion coefficient is simply
multiplied by a factor which accounts for the position-dependent velocity.
Hence the expression (\ref{fDiff}).

Plugging Eq. (\ref{fDiff}) into (\ref{mobilitydiffusion}), we obtain the
expression of the mobility,
\begin{equation}
  \mathcal{M}(X) = - \frac{d - 2}{2} \frac{\mathcal{D}_0}{E\sqrt{1 +
      \frac{X}{E}}}\,. 
  \label{mobility}
\end{equation}

Remarkably, the mobility coefficient vanishes for a two-dimensional
billiard. In this case, the Fokker-Planck equation (\ref{bFPeq}) therefore
simplifies to
\begin{equation}
  \partial_t \mathcal{P}(X,t) = \partial_X [\mathcal{D}(X) 
  \partial_X \mathcal{P}(X,t)]\,.
\label{fFPeq}
\end{equation} 
An equivalent equation was derived by Chernov and
Dolgopyat in \cite{CD07a}. This is a diffusive equation \emph{without} a
drift and describes the recurrent motion of the two-dimensional Galton
board trajectory at the macroscopic scale. In contrast, we notice that the
Fokker-Planck equation (\ref{bFPeq}) associated to a three-dimensional
version of the conservative Galton board has a non-vanishing mobility
coefficient (\ref{mobility}) and therefore retains a drift term. 

In the sequel we will assume $E>0$ so as to avoid the
singularities that come with zero velocity trajectories.\

We notice, on the one hand, that reflection at the boundaries (RBC) induces
an equilibrium state of Eq.~(\ref{fFPeq}) with constant density, 
\begin{equation}
  \mathcal{P}(X) = 1,\quad (\mathrm{RBC})\,.
  \label{fFPRBC}
\end{equation}
Flux boundary conditions (FBC), on the other hand, \emph{viz.}
\begin{equation}
  \left\{
    \begin{array}{c}
      \mathcal{P}(0) \equiv \mathcal{P}_-,\\
      \mathcal{P}(L) \equiv \mathcal{P}_+,
    \end{array}
  \right.
  \label{gbPFBC}
\end{equation}
admit the stationary state 
\begin{equation}
  \mathcal{P}(X) = C_0 + C_1 \sqrt{1 + X/E},
  \quad (\mathrm{FBC})\,.
  \label{fFPss}
\end{equation}
The coefficients $C_0$ and $C_1$ are determined by the boundary
conditions at $X=0$ and $X=L$, Eq.~(\ref{gbPFBC}):
\begin{eqnarray}
  C_0 &=& \frac{\mathcal{P}_- \sqrt{1+ L/E}- \mathcal{P}_+ }
  {\sqrt{1+ L/E} - 1},\\
  C_1 &=& \frac{\mathcal{P}_+ - \mathcal{P}_-}
  {\sqrt{1+ L/E} - 1}.
  \label{WBCcoeffs}
\end{eqnarray}
In terms of $\mathcal{P}_\pm$, we can rewrite Eq.~(\ref{fFPss}) as 
\begin{equation}
  \mathcal{P}(X) = \mathcal{P}_- + (\mathcal{P}_+ - \mathcal{P}_-)
  \frac{\sqrt{E + X} - \sqrt{E}}{\sqrt{E + L} - \sqrt{E}}\,.
  \label{fFPss2}
\end{equation}

Given rates $\mathcal{P}_-\neq\mathcal{P}_+$, the current associated to
the non-equilibrium stationary state is constant and, according to Fick's
law, equal to   
\begin{eqnarray}
  \mathcal{J} &=& - \mathcal{D}(X)
  \partial_X \mathcal{P}(X)\,,\nonumber\\
  &=& -\frac{\mathcal{D}_0}{2 E}\frac{\mathcal{P}_+ - \mathcal{P}_-}
  {\sqrt{1+L/E} - 1}\,.
  \label{masscurrent}
\end{eqnarray}
The corresponding local rate of entropy production is given according 
to the usual formula, by the product of the mass current
(\ref{masscurrent}) and the associated thermodynamic force \cite{GND03},
\begin{eqnarray}
  \frac{d_\mathrm{i}\mathcal{S}(X)}{dt} &=& \mathcal{D}(X)\frac{[\partial_X
    \mathcal{P}(X)]^2} {\mathcal{P}(X)}.
  \label{gbFPep}
\end{eqnarray}

\subsection{\label{subsec.dzd}Discretized Process}

The deterministic models we consider are to be analyzed in terms of return
maps, which involves a discretization of both time and length
scales. We consider this problem in some detail, as it will be useful for
the sake of defining a discrete process associated to the Galton board.

\begin{widetext}
Let the discretized time and length scales be determined according to $t =
k\tau$ and $X = nl$. Collision rates are proportional to the velocity,
which brings in a factor $\sqrt{1 +  nl/E}$ after we time discretize
Eq.~(\ref{fFPeq}),
\begin{eqnarray}
\lefteqn{\frac{1}{\tau}\sqrt{ 1+ \frac{n l}{E}}
\Big[\mathcal{P}(n l, k \tau + \tau) - \mathcal{P}(n l, k\tau)\Big]} 
\label{FPdisc}\\
&=&\frac{1}{l^2}
\Big\{
\mathcal{D}(n l + l/2) \mathcal{P}(n l + l, k \tau) + 
\mathcal{D}(n l - l/2) \mathcal{P}(n l - l, k \tau) - 
[\mathcal{D}(n l + l/2) + \mathcal{D}(n l - l/2)]
\mathcal{P}(nl, k \tau)
\Big\}\,.
\nonumber
\end{eqnarray}
%\end{widetext}

We let 
\begin{equation}
  \mu_n(k)\equiv \sqrt{1 + \frac{n l}{E}}\mathcal{P}(n l, k \tau)  
  \label{rhoP}
\end{equation}
be the collision frequency on the Poincar\'e surface at
position $X = nl$, and introduce a diffusion coefficient associated to the
discrete process, 
\begin{equation}
D(n) \equiv \frac{\tau}{l^2} \mathcal{D}(nl)\,.
\label{discdiffcoeff}
\end{equation}
It is convenient to set $E \equiv (2n_0+1)l/2$, for some positive integer
$n_0$. 

Equation (\ref{FPdisc}) thus transposes to the evolution 
%\begin{widetext}
\begin{equation}
  \mu_n(k + 1) = 
  \left[1- \frac{D(n + 1/2)}{\sqrt{1 + \frac{2 n}{2 n_0 + 1}}} - 
    \frac{D(n - 1/2)}{\sqrt{1 + \frac{2 n}{2 n_0 + 1}}}\right]\mu_n(k) +
  \frac{D(n + 1/2)}{\sqrt{1 + \frac{2(n+1)}{2 n_0 + 1}}}\mu_{n + 1}(k) +
  \frac{D(n - 1/2)}{\sqrt{1 + \frac{2(n-1)}{2 n_0 + 1}}}\mu_{n - 1}(k)\,.
  \label{FrPeq}
\end{equation}
\end{widetext}
Written under the form 
\begin{equation}
  \mu_n(k + 1) = \snp{n-1}\mu_{n-1}(k) + \snz{n} \mu_n(k) + \snm{n+1}
  \mu_{n+1}(k)\,, 
  \label{FrPeqrates}
\end{equation}
Eq. (\ref{FrPeq}) is seen to be the Frobenius-Perron equation of the Markov
process 
\begin{equation}
  n \longrightarrow\left\{
    \begin{array}{c@{\quad}c@{\:}c}
      n - 1,&&s_{n}^-\,,\\
      n,& \mathrm{with\ probability}\, &s_{n}^0\,,\\
      n + 1,&&s_{n}^+\,.
    \end{array}
  \right.
  \label{rw}
\end{equation}
As opposed to a symmetric random walk, the probabilities $s_n^-$,
$s_n^0$ and $s_n^+$ are asymmetric and depend on the site index,
\begin{equation}
  \begin{array}{lcl}
    \snm{n} &=& \frac{D(n - 1/2)}{\sqrt{1 + \frac{2 n}{2 n_0 + 1}}}\,,\\
    \snp{n} &=& \frac{D(n + 1/2)}{\sqrt{1 + \frac{2 n}{2 n_0 + 1}}}\,,\\
    \snz{n} &=& 1 - \snm{n} - \snp{n}\,.
  \end{array}
  \label{rwrates}
\end{equation}
In these expressions, $n$ is assumed to be a positive integer, $0\leq n \leq
N$. From Eq.~(\ref{fDiff}), the diffusion coefficient may be written
$D(n) = D_0 \sqrt{1 + \frac{2 n}{2 n_0 + 1}}$, where $D_0 = \tau/l^2
\mathcal{D}_0$, from which it follows that
\begin{equation}
  s_n^\pm = D_0\sqrt{1\pm \frac{1}{2 (n_0+ n) + 1}}\,.
  \label{rsn}
\end{equation}

It is straightforward to check that the stationary state of
Eq.~(\ref{FrPeqrates}) is independent of $D_0$, and can be
written under the form
\begin{equation}
  \mu_n \equiv\lim_{k\to\infty}\mu_n(k) = \sqrt{\frac{2 (n_0 + n) + 1} 
    {2 n_0 + 1}} P_n,
\label{ssFrPeq}
\end{equation}
where $P_n$ is the discretized stationary state of the Fokker-Planck
equation (\ref{fFPeq}),
\begin{eqnarray}
  \lefteqn{P_n \Big[\sqrt{n_0 + n + 1} + \sqrt{n_0 + n}\Big]}
  \label{FPdisc2}\\
  &&= P_{n+1} \sqrt{n_0 + n + 1} + P_{n-1}\sqrt{n_0 + n}\,.\nonumber
\end{eqnarray}
We note that the latter equation implies that $\sqrt{n_0  + n}(P_n - P_{n -
  1}) \equiv \alpha$ is constant. We can therefore write
\begin{eqnarray}
  P_n &=& P_{n-1} + \frac{\alpha}{\sqrt{n_0 + n}}\,,\nonumber\\
  &=& P_0 + \alpha \sum_{i=0}^n (n_0 + i)^{-1/2},\nonumber\\
  &=& P_0 + \alpha \Big(H_{n+n_0}^{\frac{1}{2}} - 
  H_{n_0}^{\frac{1}{2}}\Big)\,,
  \label{solP}
\end{eqnarray}
where $H_{n+n_0}^{\frac{1}{2}}$ denotes the Harmonic number,
$H_n^{\frac{1}{2}} = \sum_{j=1}^n j^{-1/2}$ \cite{Hnum}. Letting $n=N$ in
Eq.~(\ref{solP}), $Nl=L$, and 
writing the boundary 
conditions $P_0 \equiv P_-$ and $P_{N} \equiv P_+$, we obtain the
expression of $\alpha$, $\alpha = (P_+ - P_-)/(H_{N+n_0}^{\frac{1}{2}} -
H_{n_0}^{\frac{1}{2}})$. Therefore $P_n$ can be expressed as 
\begin{equation}
P_n = P_- + (P_+ - P_-)\frac{H_{n+n_0}^{\frac{1}{2}} - H_{n_0}^{\frac{1}{2}}}
{H_{N+n_0}^{\frac{1}{2}} - H_{n_0}^{\frac{1}{2}}}.
\label{solPH}
\end{equation}

The connection to the continuous case and, in particular, to
Eq.~(\ref{fFPss2}) is now straightforward. Indeed, the ratio of differences
of Harmonic functions become integrals when $l \to 0$
\begin{eqnarray}
  \lefteqn{\frac{H_{n+n_0}^{\frac{1}{2}} - H_{n_0}^{\frac{1}{2}}}
  {H_{N+n_0}^{\frac{1}{2}} - H_{n_0}^{\frac{1}{2}}}
  }
  \\
  &=& \frac{
    \frac{l}{\sqrt{E + l/2}} + \frac{l}{\sqrt{E + 3l/2}} + \dots + 
    \frac{l}{\sqrt{E + l/2 + n l}}}
  {\frac{l}{\sqrt{E + l/2}} + \frac{l}{\sqrt{E + 3l/2}} + \dots + 
    \frac{l}{\sqrt{E + l/2 + N l}}}\,,\nonumber\\
  &\to&
  \frac{\int_E^{E + X} dx/\sqrt{x}}{\int_E^{E + L} dx/\sqrt{x}} 
  = \frac{\sqrt{E + X} - \sqrt{E}}{\sqrt{E + L} - \sqrt{E}}\,,
  \nonumber
\end{eqnarray}
where the limit assumes $l\to0$ with $E$ constant and thus $n_0 \gg 1$. In
this case we have $P_n\to\mathcal{P}(X=nl)$.  

\subsection{\label{subsec.isl}Elliptic Islands}

\begin{figure}[thb]
  \centering
  \includegraphics[angle=0,width=.25\textwidth]{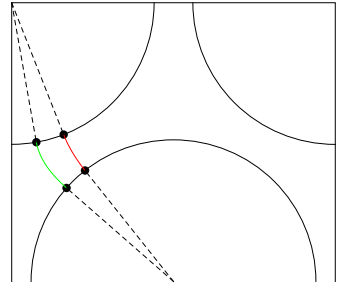}
  \caption{(Color online) The external field induces a bifurcation such that the simple
    periodic orbit bouncing off two neighboring disks at normal angles is
    replaced by two such orbits. In this situation where $E=0$, which
    corresponds to a vanishing kinetic energy on the left border, one of
    these two periodic orbits is 
    stable (to the left) and the other one unstable (to the right). The
    stability of the periodic orbit is quickly lost as $E$ is increased.}
  \label{fig.gbpo}
\end{figure}

\begin{figure}[thb]
  \centering
  \includegraphics[width=.45\textwidth]{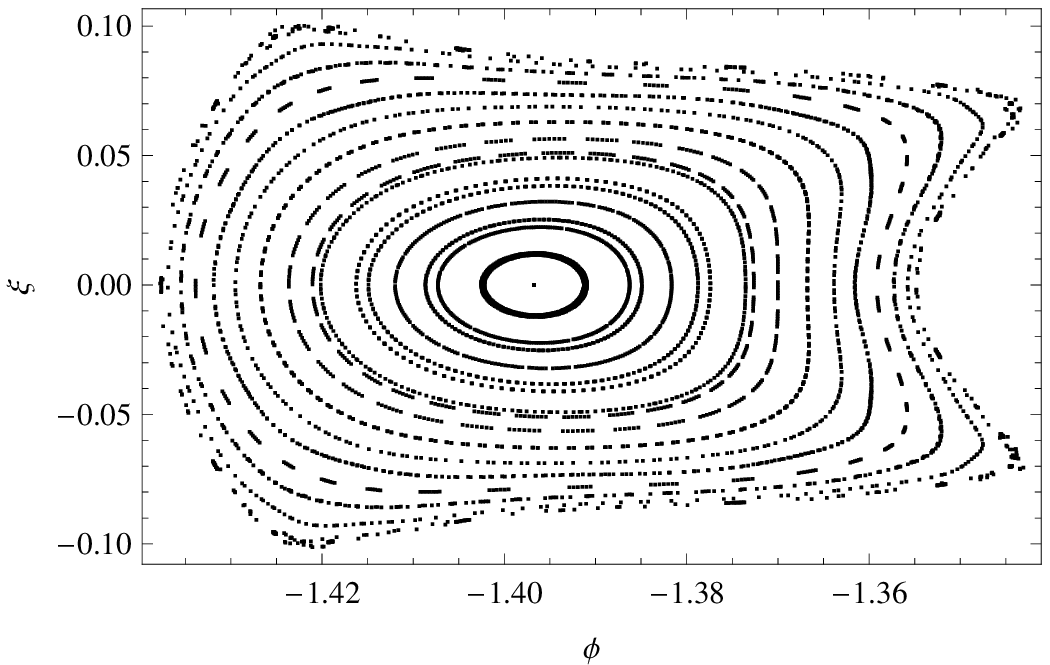}
  \includegraphics[width=.45\textwidth]{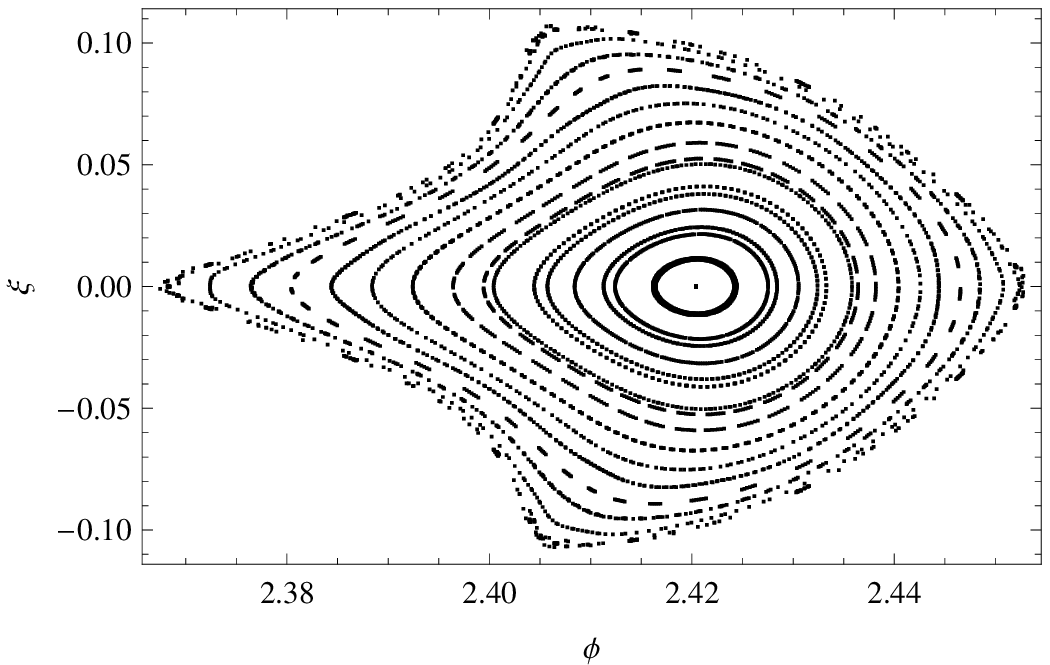}
  \caption{
    Elliptic islands around the stable periodic orbit
    shown in 
    Fig.~\ref{fig.gbpo}: (Top) disk on the upper left corner; (Bottom)
    central disk. The phase-space coordinates used here (not the
    appropriate Birkhoff coordinates) are $\phi$, the angle around the
    corresponding disk, and $\xi$ the sinus of the velocity angle measures
    with respect to the normal to the disk.} 
  \label{fig.gbpo2}
\end{figure}

Prior to turning to the stationary states of Galton boards, whether
equilibrium or non-equilibrium, we mention the possible lack 
of ergodicity of the Galton board. The external field can indeed
stabilize periodic orbits when the kinetic energy is not too large. Figure
\ref{fig.gbpo} shows such an example. In this case, elliptic islands  
co-exist with chaotic trajectories, as seen in Fig.~\ref{fig.gbpo2}.

We notice that a mixed phase space is typically expected in Hamiltonian
chaotic systems --as is the case \emph{e.g.} with the sine-circle map.
This is an undesirable feature for our own sake. However the elliptic
islands disappear if the energy value $E$ is large enough. As it turns out
of our numerical computations, $E=l/2$ is already large 
enough. We will thus assume in the sequel that $E$ large enough so the
system is fully hyperbolic. 

\subsection{\label{subsec.egb}Equilibrium Galton Board}

It is perhaps not widely appreciated that one can obtain an equilibrium
state consistent with the presence of the external field. The reason for
this is actually quite simple. Liouville's theorem implies the conservation
of the volume measure,  
\begin{eqnarray}
  d\Gamma &=& dx\,dy\,dv_x\,dv_y,\nonumber\\
  &=& v\,dE\,dt\,d\phi\,d\xi,
  \label{gbdliouville}
\end{eqnarray}
where $\phi$ and $\xi$ are defined to be the
angle along the disk and sinus of the outgoing velocity angle measured with
respect to the normal to the disk. 

We remark that because of the factor $v$ that multiplies the volume
measure in Eq.~(\ref{gbdliouville}), the pair $(\phi,\xi)$ are not
canonical variables anymore. Indeed the position along the cylinder axis
varies with the angle coordinate $\phi$, so that the velocity $v$ depends
on $\phi$. The appropriate generalized angle variable conjugated to $\xi$
can be determined accordingly \cite{Kog01}. 

\begin{figure}[htb]
  \centering
  \includegraphics[width=.45\textwidth]{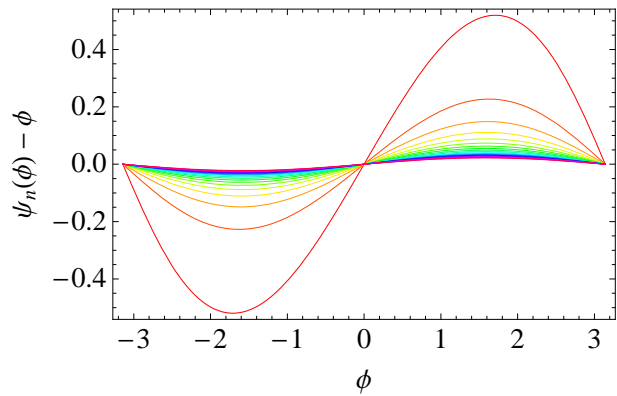}
  \caption{(Color online) Difference between the generalized angle
    coordinate $\psi_n(\phi)$ and $\phi$, here computed for $\sigma = 0.44
    l$, $n_0 = 0$, and $n=1,\dots,20$. Larger differences occur at smaller
    $n$, where the effect of the external field is most noticeable.}
  \label{fig.birk}
\end{figure}

Introducing the index $n$, referring to the $n$th disk, whose
center has position  $x = (n-1)l/2$ along the cylinder axis, the velocity at
angle $\phi$ along disk $n$ is
\begin{eqnarray}
  v_n(\phi) &=& \sqrt{2[E + (n-1)l/2 + \sigma \cos\phi]},\nonumber\\
  &=& \sqrt{(n_0 + n)l + 2 \sigma \cos\phi},
  \label{gbdpn}
\end{eqnarray}
The canonical coordinate conjugated to $\xi$ is
therefore $\psi_n$, such that $d\psi_n = v_n(\phi)\,d\phi$,
\begin{eqnarray}
  \psi_n(\phi) &=& 2\pi \frac{\int_0^\phi \, v_n(\phi)\,d\phi}
  {\int_0^{2\pi} \, v_n(\phi)\,d\phi}\,,
  \nonumber\\
  &=& \pi \frac{\mathrm{E}
    \left(\frac {\phi} {2}, 2\frac {2\sigma} {(n + n_0)l +
        2\sigma}\right)}
  {\mathrm{E}\left(2\frac{2\sigma}{(n + n_0)l + 2\sigma}\right)}\,,
  \label{gbdbirkhoff}
\end{eqnarray}
where $\mathrm{E}$ denotes the elliptic integral of the second kind, 
$\mathrm{E}(\phi,x) = \int_0^{\phi/2}\sqrt{1-x\sin^2\theta}d\theta$, and
$\mathrm{E}(x) = \mathrm{E}(\pi/2,x)$ is the complete elliptic integral. As
seen in 
Fig. \ref{fig.birk}, the difference between $\psi_n$ and $\phi$ decreases
rapidly as $nl$ increases. Note that $\sigma$ is assumed to scale with $l$
so that $\psi_n$ does not actually depend on $l$.

\begin{figure}[htb!]
  \centering
  \includegraphics[angle=0,width=.38\textwidth]{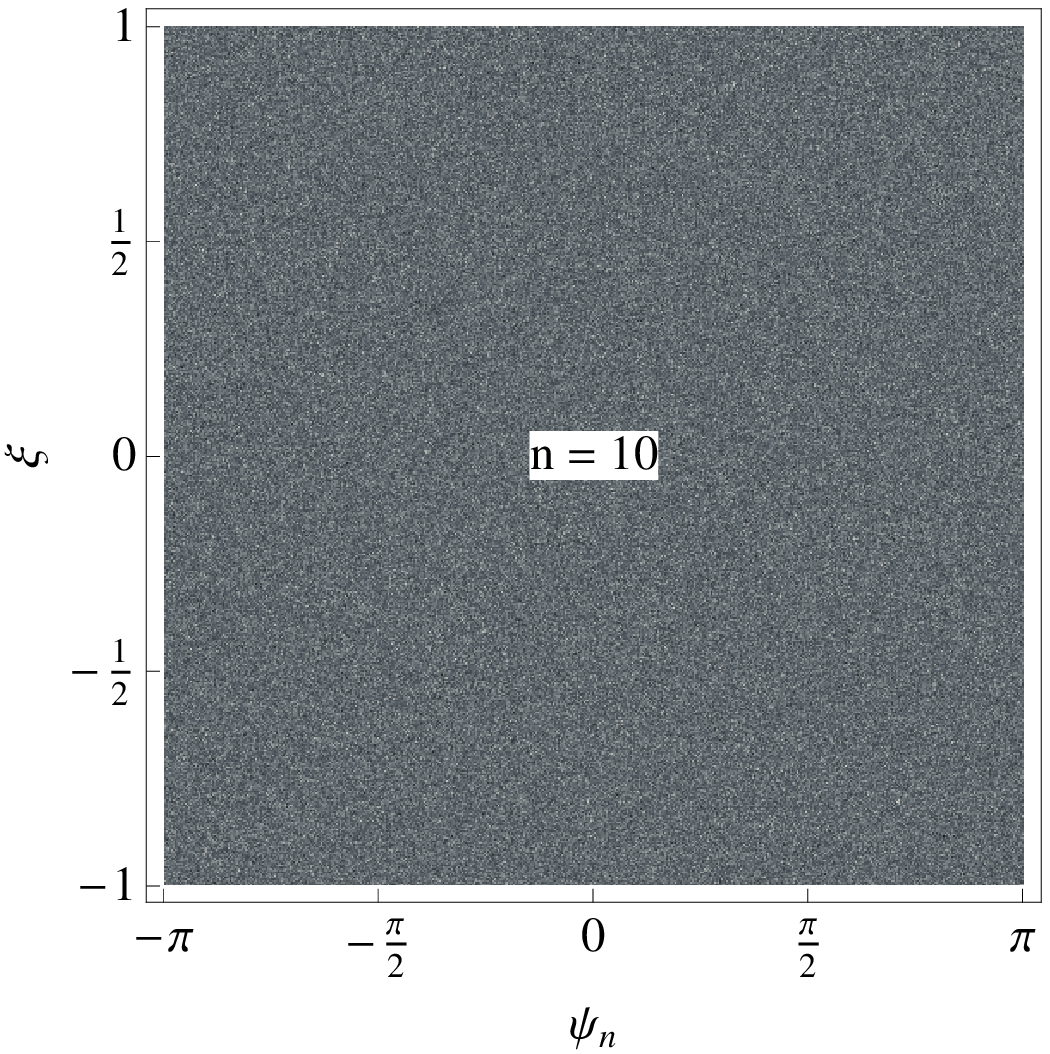}
  %\hfill
  \includegraphics[angle=0,width=.38\textwidth]{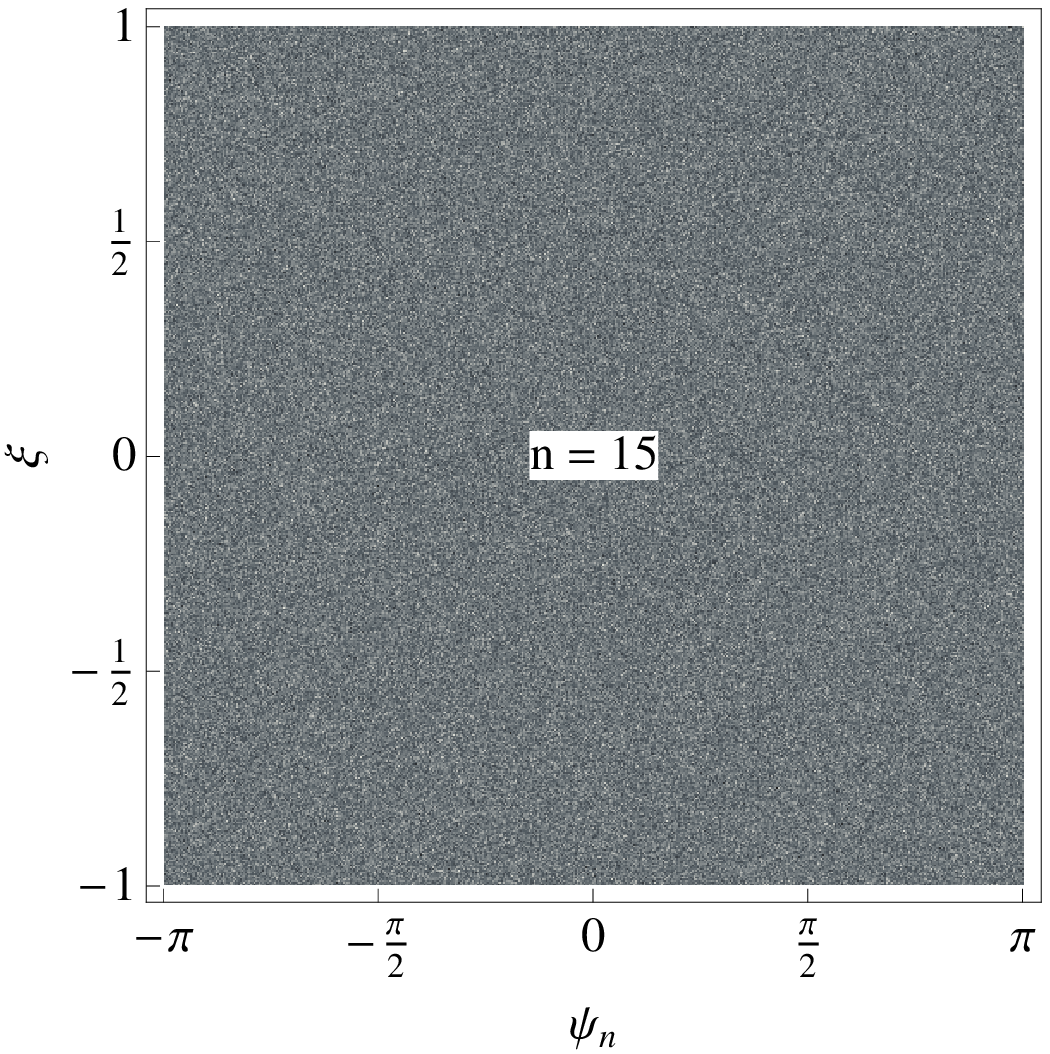}
  %\hfill
  \includegraphics[angle=0,width=.38\textwidth]{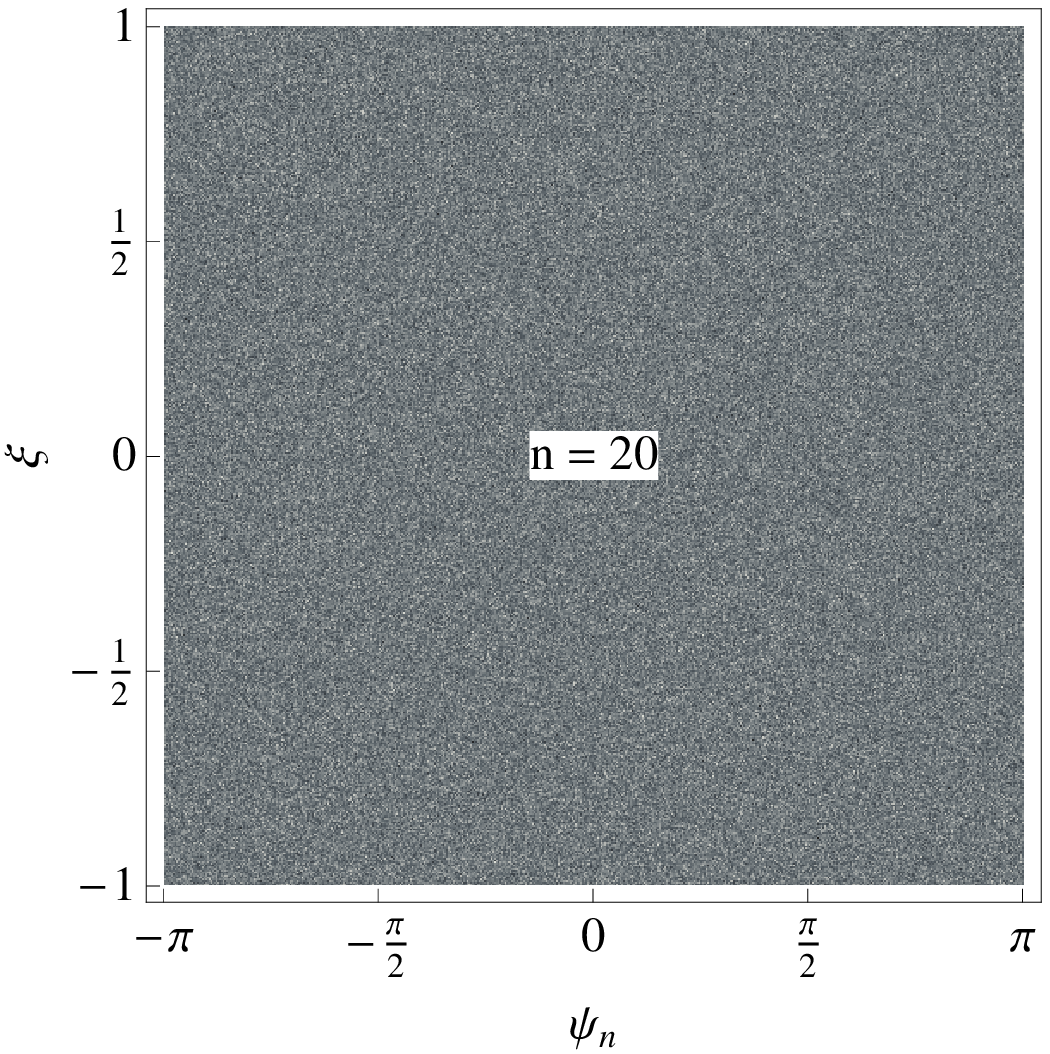}
  \caption{Invariant density associated to a closed Galton board of size
    $L=10l$, with reflection at the boundaries $x=0$ and $x=10l$, and energy
    $E=l/2$. This is an equilibrium system as reflected by the uniformity
    of the phase portraits.} 
  \label{fig.cgbpp}
\end{figure}

Let us consider a closed Galton board of length $L = N l$ ($2N + 1$ disks)\,,
with reflecting boundaries at $x=0$ and $x = L$. This is an equilibrium
system. More precisely the invariant density associated to each disk is
uniform, as verified in Fig.~\ref{fig.cgbpp}. The distinctive feature
however is that the time scale changes with the disk index $n$, $\tau(n)
\sim 1/v_n$. Thus particles move faster with increasing $n$, but
correspondingly they make more collisions so that their distribution is
uniform in time. 

\begin{figure}[thb]
  \centering
  \includegraphics[width=.45\textwidth]{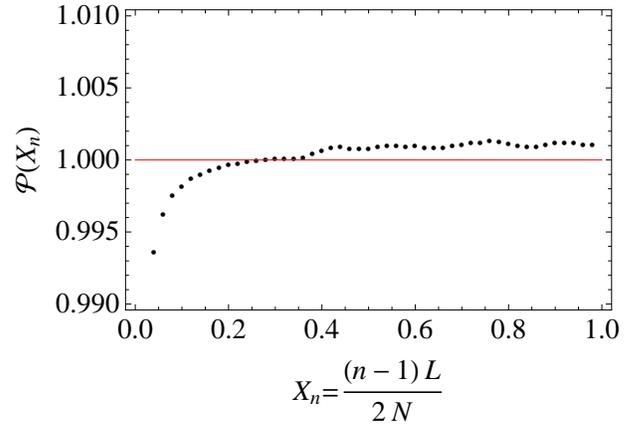}
  \caption{Equilibrium stationary density of the closed Galton board
    obtained for a channel of length $L = 1$, with $2N+1$ disks,
    $N=25$. The solid line is the constant equilibrium density
    $\mathcal{P}(X) = 1$.}
  \label{fig.cgbss}
\end{figure}

From the average count of collision events of disk $n$, we obtain the
collision frequency, which, when multiplied by the local time scale (this
amounts to dividing it by the velocity $v_n$ evaluated at the center of
cell $n$) yields the average density $P_n \sim \mathcal{P}(X_n =
(n-1)l/2)$. This quantity, shown in Fig.~\ref{fig.cgbss}, is indeed found
to be almost constant, thus confirming our reasoning.

\subsection{\label{subsec.ngb}Non-Equilibrium Galton Board}

A non-equilibrium stationary state of the Galton board can be achieved much
in the same way as with the open Lorentz gas studied in \cite{BGGi}, by
assuming that a flux of trajectories is continuously flowing through the
boundaries which are let in contact with stochastic particle reservoirs at
$x = 0$ and $x = L$.
\begin{equation}
  \rho(\Gamma, t)\Big|_{x=0, L} = \rho_{\pm}.
  \label{gbrhobc}
\end{equation}

In analogy to the field free case, the invariant solution of the Liouville
equation compatible with the boundary conditions (\ref{gbrhobc}) is given,
for almost every phase point $\Gamma$, by
\begin{eqnarray}
  \rho(\Gamma)
    &=& \rho_- + \frac{\rho_+-\rho_-}{\sqrt{2(E+L)} - \sqrt{2E}} 
  \label{gbss1}\\
  &&\times
  \left[v(\Gamma) - \sqrt{2E} + 
    \int_0^{-T(\Gamma)} dt \dot v(\Phi^t\Gamma)\right]\,.
  \nonumber
\end{eqnarray}
Here $\rho(\Gamma)$ is written in terms of the change in velocity
amplitude, given by $v(\Gamma) = \sqrt{2[E + x(\Gamma)]}$ at the
corresponding horizontal position  $x(\Gamma)$. Thus $\dot v(\Gamma)
= [\dot x(\Gamma)]/v(\Gamma)$ and, provided the change in velocity between
successive collisions is small, we can write $\int_0^\tau \dot v(\Phi^t
\Gamma) dt \simeq [x(\Phi^\tau\Gamma) - x(\Gamma)]/v(\Gamma)$. Hence,
denoting by $\tau_k$ the time separating the $(k-1)$th and $k$th collisions
and by $t_k = \sum_{j=1}^k \tau_k$ the time elapsed after $k$ collisions,
we have
\begin{eqnarray}
  \rho(\Gamma)
  &\simeq& \rho_- + \frac{\rho_+-\rho_-}{\sqrt{2(E+L)} - \sqrt{2E}} 
  \Bigg[v(\Gamma) - \sqrt{2E} 
  \nonumber\\
  && + \sum_{k=1}^{K(\Gamma)} 
    \frac{x(\Phi^{-t_k}\Gamma) - x(\Phi^{-t_{k-1}}\Gamma)}
    {v(\Phi^{-t_k}\Gamma)}
  \Bigg]\,.
\end{eqnarray}
This approximation becomes exact when the number of cells in the system is
let to infinity, in which case $K(\Gamma)$, the number of collisions for
the trajectory to reach the boundaries becomes infinite. Therefore the
invariant state is 
\begin{eqnarray}
  \rho(\Gamma)
  &=& \rho_- + \frac{\rho_+-\rho_-}{\sqrt{2(E+L)} - \sqrt{2E}} 
  \Bigg[v(\Gamma) - \sqrt{2E} 
  \nonumber\\
  && + \sum_{k=1}^{\infty}
  \frac{x(\Phi^{-t_k}\Gamma) - x(\Phi^{-t_{k-1}}\Gamma)}
  {v(\Phi^{-t_k}\Gamma)}
  \Bigg]\,.
  \label{gbinfss}
\end{eqnarray}
so that the fluctuating part of the invariant density becomes
singular. This is analogous to the field free case discussed in \cite{BGGi}.

We compute this quantity numerically from the statistics of the Birkhoff
map of the Galton board, using a cylindrical Galton board
similar to that shown in Fig.~\ref{fig.gbd}, with
external forcing of unit magnitude in the direction of the cylinder axis,
letting the particles have energy $E = 1/2$. The particles are thus
injected at $x=0$ with unit velocity at random angles and subsequently
absorbed upon their first passage to either $x=0$ or $x=L$.

\begin{figure}[thb]
  \centering
  \includegraphics[angle=0,width=.45\textwidth]{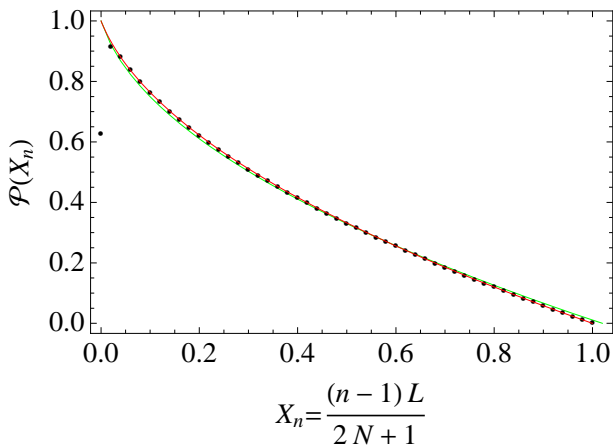}
  \caption{(Color online) Non-equilibrium stationary density of the Galton
    board obtained for a channel of length $L = 1$, with $51$ disks
    ($N=25$). Two solid lines are shown which are barely distinguishable,
    corresponding to Eqs.~(\ref{fFPss2}) (Red) and (\ref{solPH}) (Green)
    with $E = l/2$ and thus $n_0 = 0$.}
  \label{fig.ogbss}
\end{figure}

The computation of the collision frequency at disk $n$, averaged over the
phase-space coordinates yields the quantity $\mu_n$, Eq.~(\ref{ssFrPeq}),
which, after dividing by the modulus of the velocity at that site, is
converted to $P_n$, the stationary solution of the Fokker-Planck equation
(\ref{fFPeq}). Here, we have 
\begin{eqnarray}
  \mathcal{P}(X_n) &=& 
  \frac{1}{l}\int_{\mathbb{C}_n\}}d\Gamma
  \rho(\Gamma)\,,\nonumber\\
  &=& \mathcal{P}_- + (\mathcal{P}_+ - \mathcal{P}_-)
  \frac{\sqrt{(E + X_n)} - \sqrt{E}} 
  {\sqrt{(E + L)} - \sqrt{E}} \,.
  \label{ccgbP}
\end{eqnarray}
The results of this computation are presented in Fig.~\ref{fig.ogbss}, 
and compared to Eqs.~(\ref{fFPss2}) and (\ref{solPH}). The agreement with
both discrete and continuous solutions is excellent.

\begin{figure*}[tbph]
  \centering
  \includegraphics[angle=0,width=.4\textwidth]{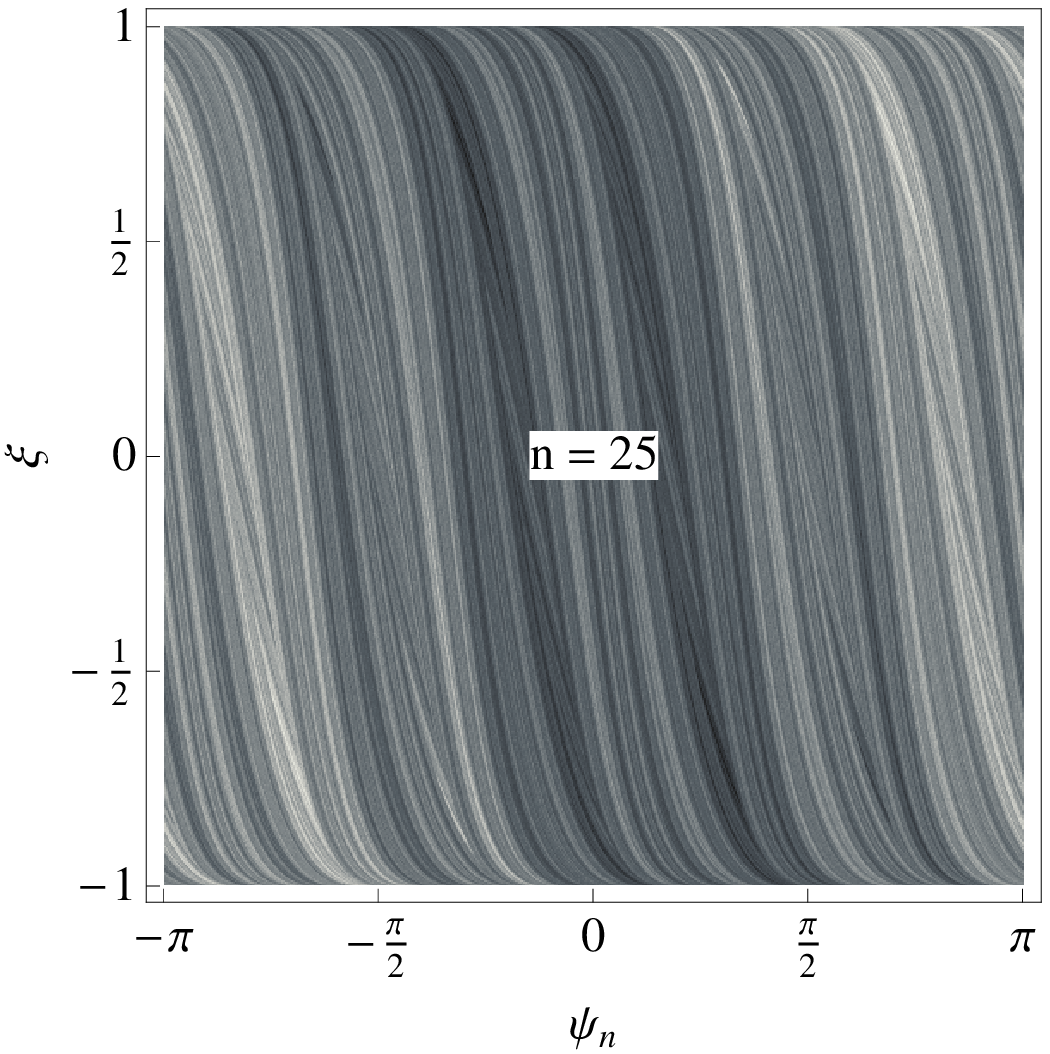}
  \hfill
  \includegraphics[angle=0,width=.4\textwidth]{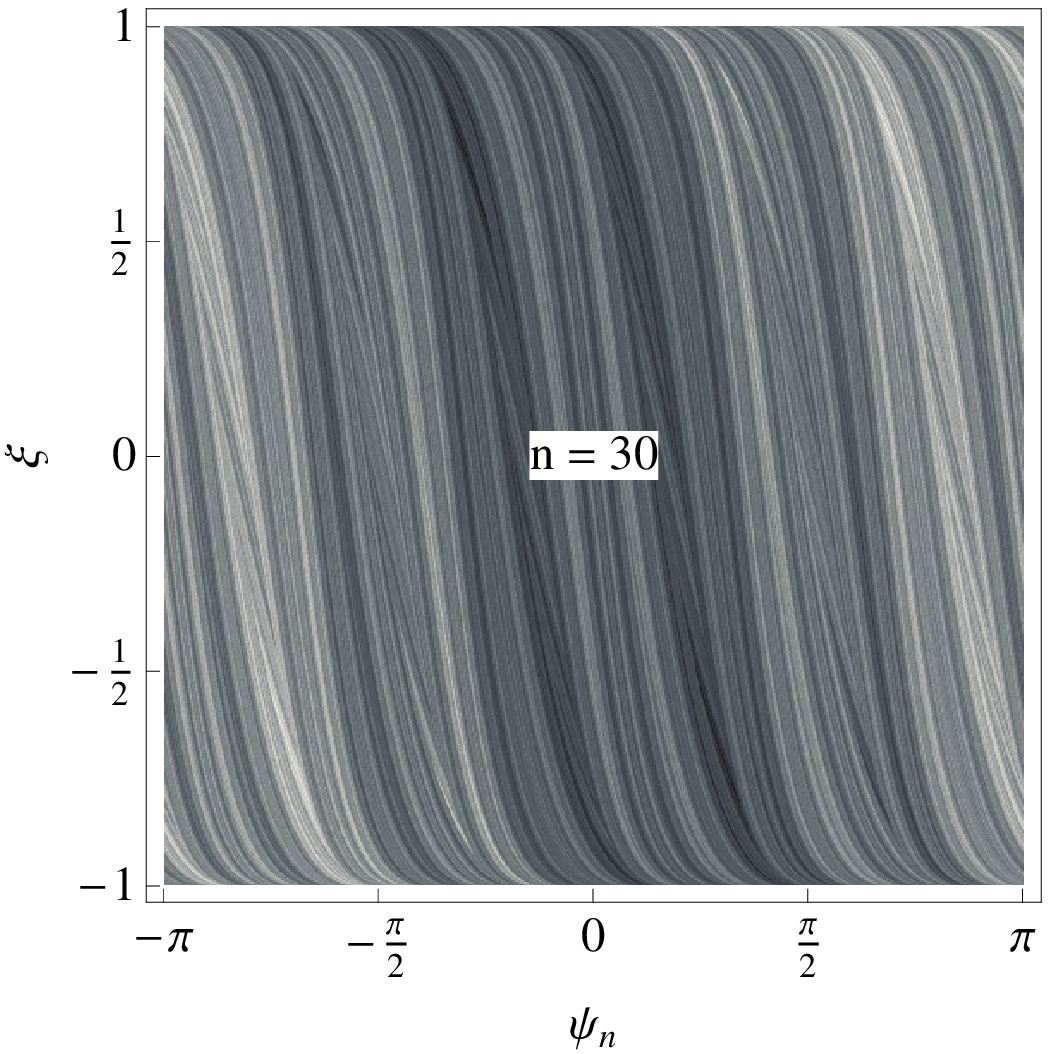}
  \includegraphics[angle=0,width=.4\textwidth]{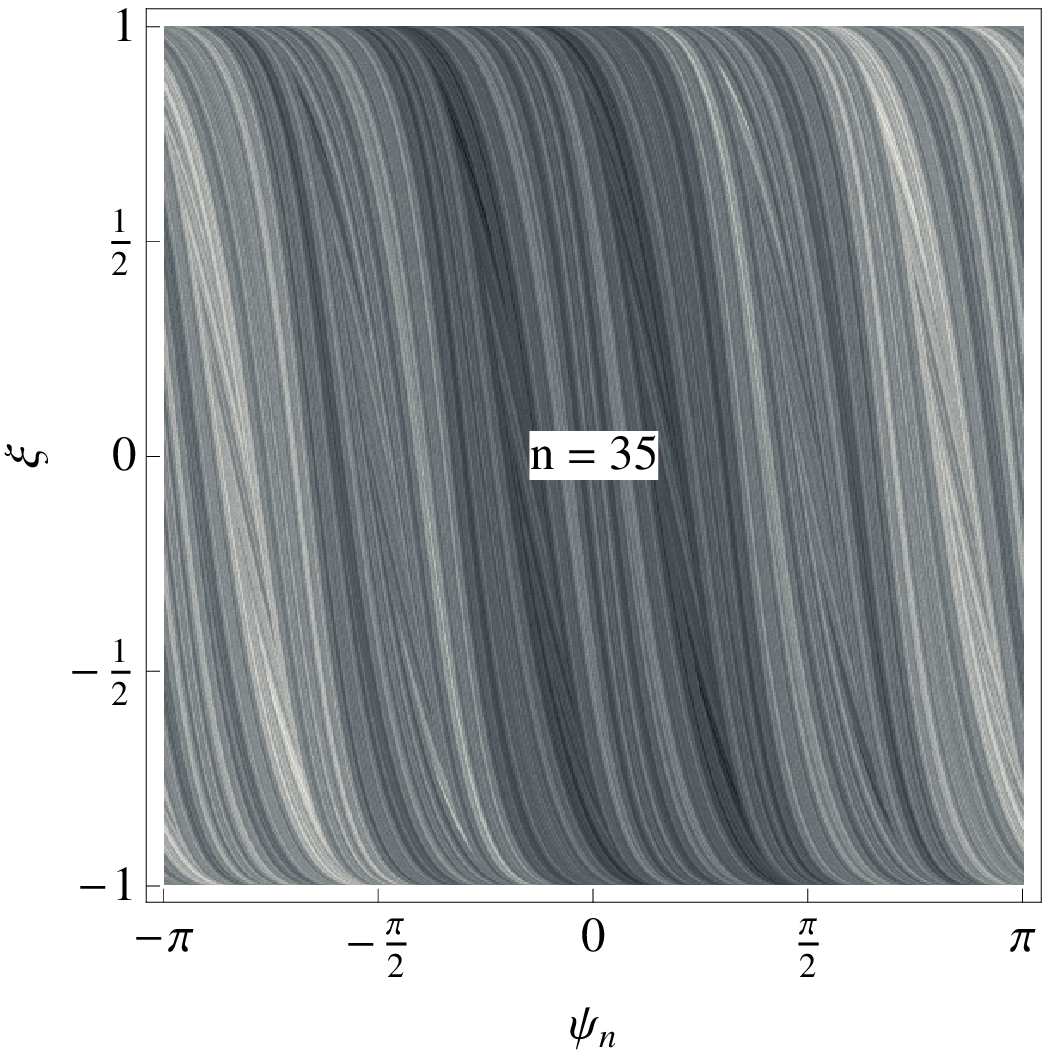}
  \hfill
  \includegraphics[angle=0,width=.4\textwidth]{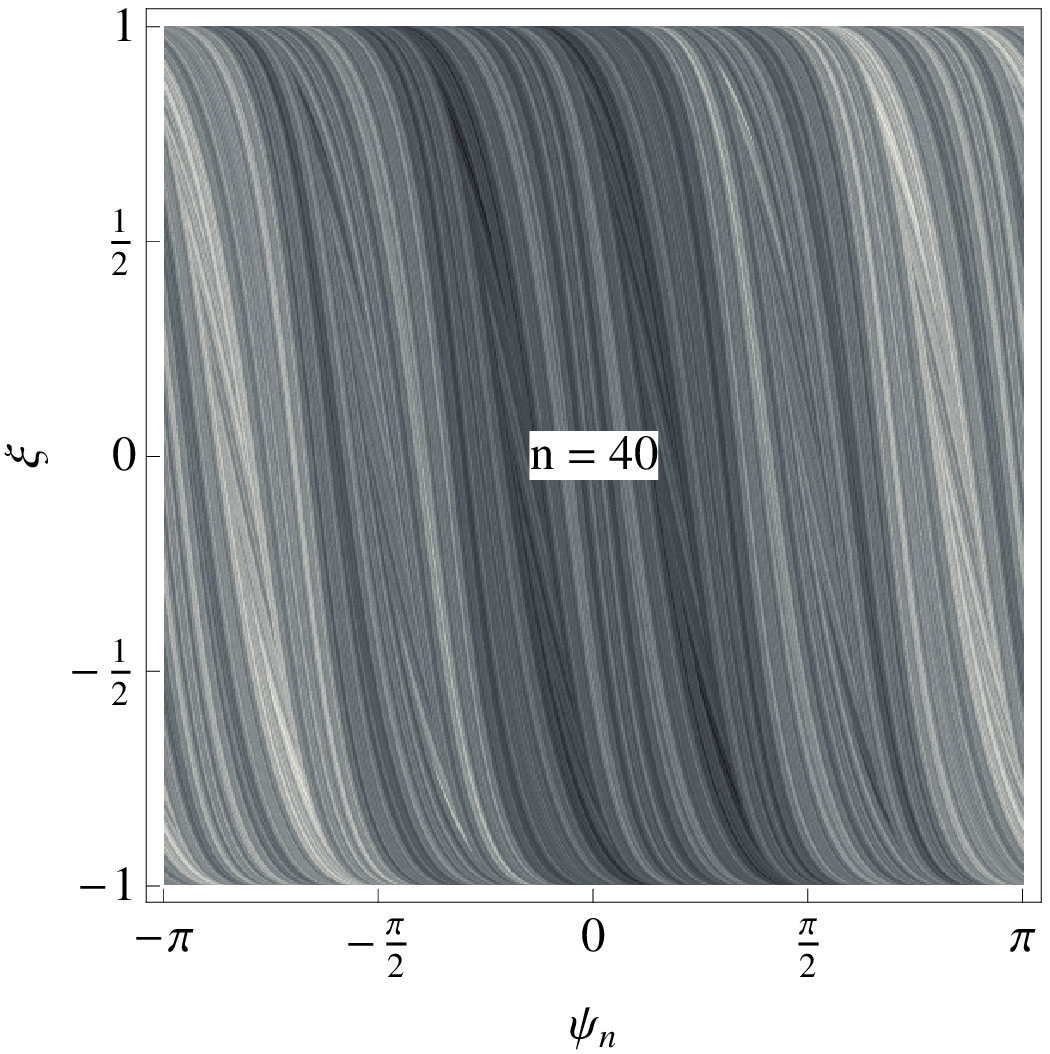}
  \includegraphics[angle=0,width=.4\textwidth]{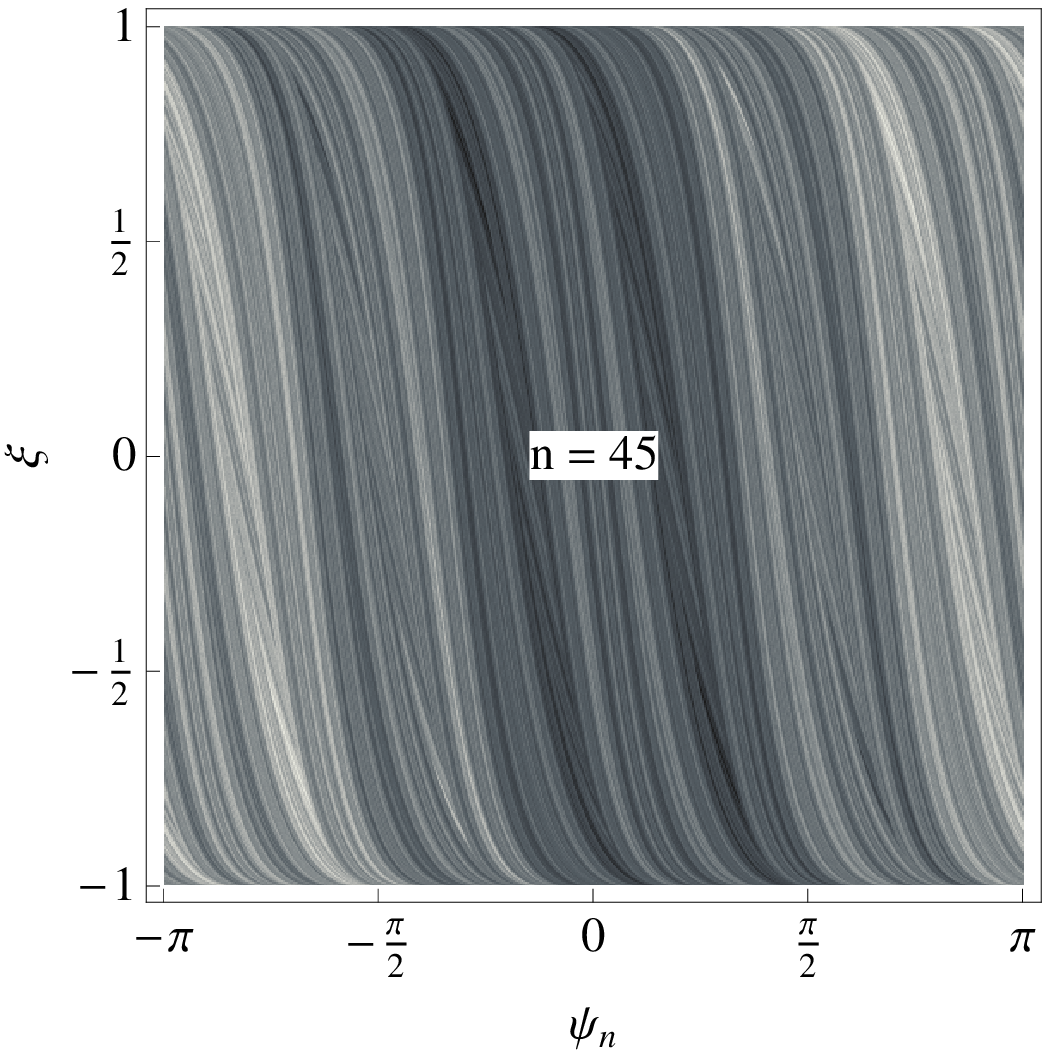}
  \hfill
  \includegraphics[angle=0,width=.4\textwidth]{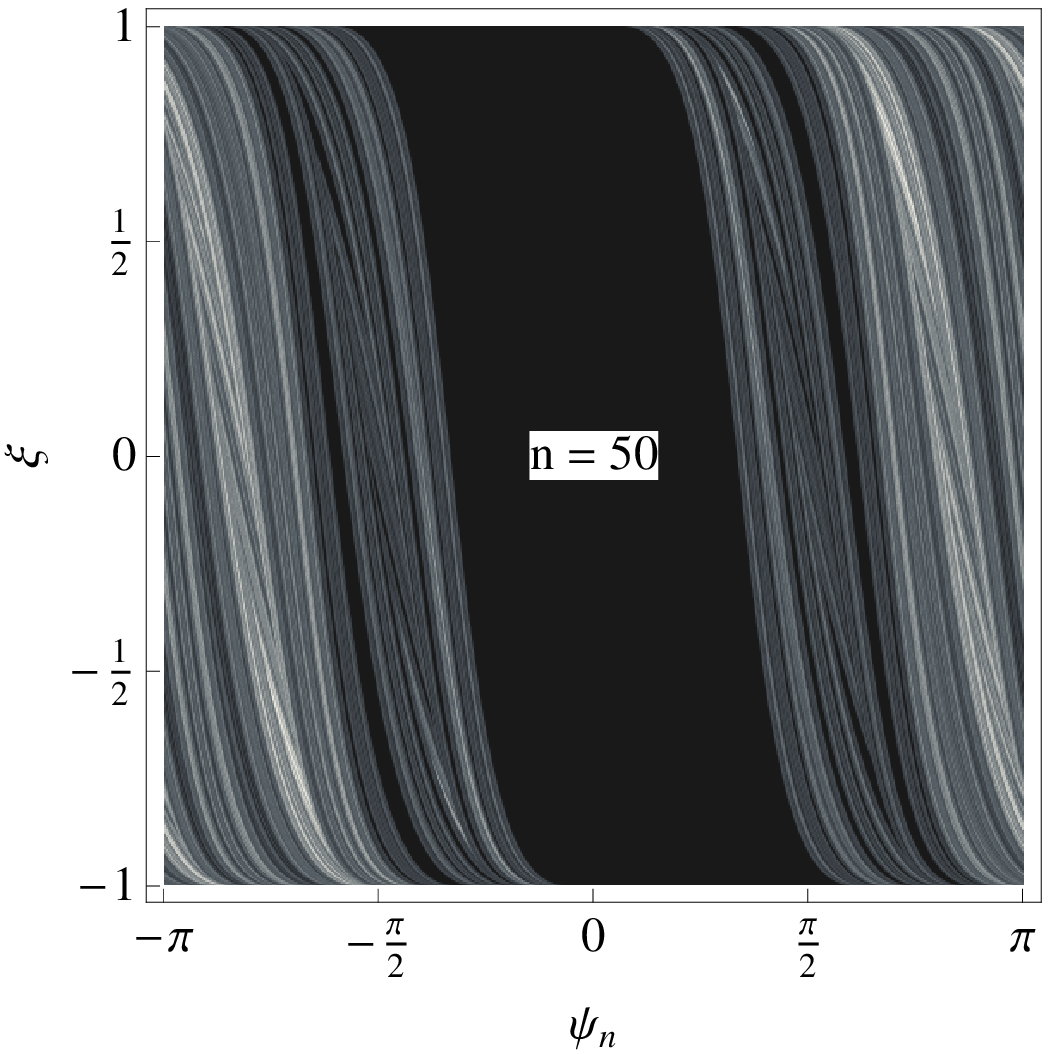}
  \caption{Non-equilibrium phase-space densities of the open Galton channel
    with a geometry similar to that shown in Fig.~\ref{fig.gbd}, with
    absorbing boundaries at $x = 0$ and $x = 1$, and stochastic injection
    of particles at $x = 0$ only. Disk $50$ is the one before last. Black
    areas correspond to absorption at the nearby boundary. The color white
    is associated to injection from the left boundary. Thus hues of gray
    correspond to phase-space regions with mixtures of phase-space points
    which are mapped backward to the left and right borders. The
    corresponding overall densities are shown in  Fig.~\ref{fig.ogbss}.  
 } 
  \label{fig.ogbpp}
\end{figure*}

The histograms displayed in Fig.~\ref{fig.ogbpp} show the fluctuating part
of the invariant phase-space density computed in terms of the Birkhoff
coordinates $(\psi_n,\xi)$, Eq.~(\ref{gbdbirkhoff}). The fractality of
these graphs is much like that of the graphs of the open Lorentz gas,
see \cite{BGGi}. The differences are indeed too tenuous to tell. As
with the closed Galton board though, the distinctive feature is that the
collision rates increase with the cell index with the amplitude of the
velocity. 

To further analyze the fractality of the stationary state of the
non-equilibrium Galton board and its relation to the phenomenological
entropy production, Eq.~(\ref{gbFPep}), we introduce in the next section an
analytically tractable model, which generalizes the multi-baker map
associated to a field-free symmetric diffusion process, so as to account
for the acceleration of tracer particles under the action of the external
forcing.  

%\begin{widetext}
\section{\label{sec.fmb}Forced multi-baker map}

A time-reversible volume-preserving deterministic process can be associated
to Eq.~(\ref{FrPeqrates}) in the form of a multi-baker map with energy,
defined on the phase space ${(n,[0,l_n]\times[0,l_n])}_{n\in\mathbb{Z}}$,
where each unit cell has area $l_n^2 \equiv a_n l^2$, $a_n = \sqrt{}[1+
  (2n)/(2n_0+1)]$, and the dynamics is defined according to
\begin{widetext}
\begin{equation}
  B:\,(n,x,y)
  \mapsto
  \left\{
    \begin{array}{l@{\quad}l}
      \left(n-1,\frac{l_{n-1}}{l_n} \frac{x}{\snm{n}}, \frac{l_{n-1}}{l_n}
        \snp{n-1}y\right)\,,&0\leq x\leq l_n \snm{n},\\ 
      \left(n,\frac{x-\snm{n}l_n}{\snz{n}},\snp{n}l_n+\snz{n}y\right)\,,
      &l_n \snm{n}\leq x \leq l_n(1-\snp{n})\,,\\ 
      \left(n+1,\frac{l_{n+1}}{l_n} \frac{x-\snm{n}l_n-\snz{n}l_n}{\snp{n}},
        \frac{l_{n+1}}{l_n} (\snp{n+1}l_{n}+\snz{n+1}l_{n}+\snm{n+1}y)
      \right)\,, 
      & l_n(1-\snp{n})\leq x\leq l_n.
    \end{array}
  \right.
  \label{fmbmap}
\end{equation}
\end{widetext}
This map has two important properties. First, the areas of the unit cells
are chosen to vary with the amplitude of the velocity, which
ensures that the Jacobian of $B$, $\frac{a_{n-1}\snp{n-1}}
{a_n \snm{n}}$ or $\frac{a_{n+1}\snm{n+1}}{a_n \snp{n}}$, is unity. Second,
$B$ is time-reversal symmetric under the operator $S:\,(n,x,y) \to
(n,l_n-y,l_n-x)$, {\em i.e.} $S\circ B = B^{-1}\circ S$, as is easily
checked.
\begin{figure}[htb]
  \centering
  \includegraphics[angle=0,width=.45\textwidth]{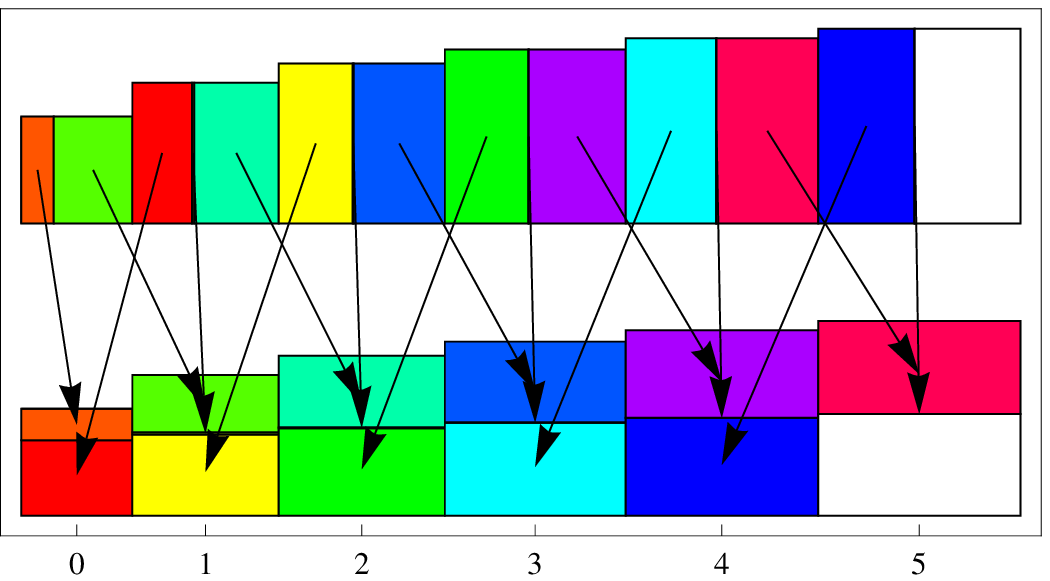}
  \caption{(Color online) Forced mutli-baker map (\ref{fmbmap}). The cells
    have areas $a_n l^2$, with coordinates $(x,y)$, and  labeled by a
    positive integer $n$ ($n_0=0$ here)\,. The map divides each cell into
    three vertical 
    rectangles. The left rectangle is mapped to the bottom horizontal
    rectangle in the left neighboring cell. Likewise the right rectangle is
    mapped to the top horizontal rectangle in the right neighboring cell. The
    middle vertical rectangle is mapped to the middle horizontal rectangle
    in the same cell, but are so tiny they are barely visible already for
    $n=1$.} 
  \label{fig.fmbmap}
\end{figure}

Multi-baker maps with energy have been considered earlier
\cite{TG99,TG00}. The novelty here is to introduce $n$-dependent rates
$s_n^\pm$ and $s_n^0$, Eq.~(\ref{rsn}). We will assume $D_0=1/2$ in the
sequel, so that, provided $n_0 + n \gg 1/2$, we can write 
\begin{equation}
  s_n^\pm = \frac{1}{2} \pm \frac{1}{4}\frac{1}{2 (n_0+ n) + 1}
  - \frac{1}{16}\frac{1}{[2 (n_0+ n) + 1]^2} + \dots
  \label{rsnexpand}
\end{equation}
Thus $\snz{n} = 1 - \snp{n} - \snm{n}$ is approximated by
\begin{equation}
  \snz{n} = \frac{1}{8}\frac{1}{[2 (n_0+ n) + 1]^2} + \dots
  \label{rsnzexpand}
\end{equation}
which is vanishingly small. Therefore, when $n+n_0$ is large, the dynamics
of $B$ reduces to that of the usual multi-baker map, at the exception
of the energy dependence which fixes the local time scales.

\subsection{\label{subsec.ste}Statistical ensembles}

An initial density of points $\Gamma = (n,x,y)$, $\rho(\Gamma,0)$, evolves
under repeated iterations of $B$ according to the action of the
Frobenius-Perron operator, which, since $B$ preserves phase-space
volumes, is simply given by $\rho(\Gamma,k+1) =  \rho(B^{-1}\Gamma,k)$.
In order to characterize the stationary density, $\rho(\Gamma) =
\lim_{k\to\infty} \rho(\Gamma,k)$, we consider the cumulative function
$\mu_n(x,y,k) = \int_0^{x} dx' \int_0^y dy' \rho(n,x',y',k)$. Notice that
$\rho$ here refers to the statistics of the return map and therefore
differs from the density $\rho$ associated to the Galton board,
Eq.~(\ref{gbinfss}), by a factor proportional to the local time scale.

The identification of this function proceeds along the
lines of Refs. \cite{TG95,TGD98,GD99}. Under the assumption that the
$x$ dependence of the initial density is trivial, we can write $\mu_n(x,y,k) =
x/\l_n \mu_n(l_n,y,k)$. Letting $0\leq y \leq 1$, it is then easy to verify
that $\mu_n(l_n,y\, l_n,k)$ obeys the functional equation
\begin{widetext}
\begin{equation}
  \mu_{n}(l_n, y\, l_n, k+1)
  =
  \left\{
    \begin{array}{l@{\quad}l}
      \snm{n+1}\mu_{n+1}\left(l_{n+1}, \frac{y}{\snp{n}} l_{n+1}, k
      \right)\,, &0\leq y\leq \snp{n},\\ 
      \snm{n+1}\mu_{n+1}(l_{n+1}, l_{n+1}, k) + 
      \snz{n}\mu_n\left(l_n, \frac{y - \snp{n}}{\snz{n}} l_n , k\right)\,,
      &\snp{n}\leq y \leq 1-\snm{n},\\ 
      \snm{n+1} \mu_{n+1}(l_{n+1}, l_{n+1}, k) + 
      \snz{n}\mu_n(l_n, l_n, k) &\\
      \quad + 
      \snp{n-1} \mu_{n-1}\left(l_{n-1}, \frac{y -
          \snp{n}-\snz{n}}{\snm{n}} l_{n-1}, k \right)\,, 
      & (1-\snm{n})\leq y\leq 1.
    \end{array}
  \right.
  \label{pfbm}
\end{equation}
\end{widetext}
In particular, letting $y = l_n$, we recover
\begin{eqnarray}
  \lefteqn{
    \mu_{n}(l_n,l_n,k+1) =
    \snm{n+1} \mu_{n+1}(l_{n+1}, l_{n+1},k)}\\
  &&+ \snz{n} \mu_n(l_n,l_n,k)
  + \snp{n-1} \mu_{n-1}(l_{n-1}, l_{n-1},k)\,,
  \nonumber
\end{eqnarray}
which is identical to Eq.~(\ref{FrPeqrates}) with $\mu_n(k)\equiv
\mu_n(l_n, l_n, k)$. Let $\mu_n$ denote the steady state of this equation,
$\mu_n = \lim_{k\to\infty} \mu_n(l_n,l_n,k)$.  

The steady state of Eq.~(\ref{pfbm}) can be written under the form, $0\leq
x,y \leq 1$,
\begin{eqnarray}
  \mu_n(x\, l_n, y\,l_n) &\equiv& \lim_{k\to\infty} \mu_n(x\, l_n, y\, l_n,
  k)\,,\nonumber\\
  &=&  x\, y\, \mu_n + 2x\,(\snm{n+1}\mu_{n+1} - \snp{n}\mu_n) F_n(y)\,,
  \nonumber\\
  &\equiv& x\,y\, \mu_n + \alpha\, x\, F_n(y)\,,
  \label{sssubs}
\end{eqnarray}
where we introduced the generalized Takagi functions $F_n$, with a
prefactor, $\alpha = 2\big(\snm{n+1}\mu_{n+1} - \snp{n}\mu_n\big)$,  which,
as in Eqs. (\ref{FPdisc2})-(\ref{solPH}) is easily seen to be independent
of $n$: 
\begin{eqnarray}
  2\big(\snm{n+1}\mu_{n+1} - \snp{n}\mu_n\big) &=& 
    \sqrt{\frac{2(n_0 + n + 1)}{2 n_0 + 1}} (P_{n+1} - P_n),\,\nonumber\\
  &=& \sqrt{\frac{2}{2 n_0 + 1}}
  \frac{P_+ - P_-}{H_{N + n_0}^{1/2} - H_{n_0}^{1/2}}
  \,,\nonumber\\
  &\to& \frac{l}{2\sqrt{E}} \frac{P_+ - P_-}
  {\sqrt{E + L} - \sqrt{E}}\,,
  \label{alpha}
\end{eqnarray}
where the limit holds when $l\to0$. Notice that the prefactor is
proportional to $l$. Thus $\alpha$ is a small parameter in that limit.

Substituting Eq. (\ref{sssubs}) into Eq.~(\ref{pfbm}), the generalized 
Takagi function $F$ is found to satisfy the functional equation
\begin{eqnarray}
  \lefteqn{F_n(y) =}
  \label{ssbmh}\\
  &&\left\{
    \begin{array}{l@{\quad}c}
      \frac{y}{2 \snp{n}} + \snm{n+1} F_{n+1}\left(\frac{y}{\snp{n}}\right)\,,
      &0\leq y < \snp{n},\\
      \frac{1}{2}
      + \snz{n}F_n\left(\frac{y - \snp{n}}{\snz{n}}\right)\,,
      &\snp{n}\leq y < 1-\snm{n},\\ 
      \frac{1 - y}{2 \snm{n}} +
      \snp{n-1} F_{n-1}\left(\frac{y - \snp{n}-\snz{n}}{\snm{n}}
      \right)\,,  
      & 1-\snm{n} < y\leq 1.
    \end{array}
  \right.
  \nonumber
\end{eqnarray}
The boundary conditions are such that the density is uniform at
$n = 0, N + 1$, implying $F_0(y) = F_{N+1}(y) = 0$. Notice that this function
reduces to the 
Takagi function in the limit $n, N\to\infty$, $n\ll N$. Indeed
$\snm{\infty}, \snp{\infty} = 1/2$, $\snz{\infty} = 0$. Therefore
Eq.~(\ref{sssubs}) is similar to the corresponding expression obtained for
the multi-baker map, see \cite{BGGi}.

\subsubsection{Generalized Takagi functions}

For the sake of plotting $F_n(y)$, it is convenient to consider
the graph of $F_n(y)$ vs. $y$ as parameterized by a real variable,
$0\leq x\leq 1$, defined so that
\begin{eqnarray}
  \lefteqn{y_n(x) =}
  \label{yx}\\
  &&\left\{
    \begin{array}{l@{\quad}c}
      \snp{n} y_{n+1}(3 x)\,,&0\leq x < 1/3,\\
      \snp{n} + \snz{n} y_n (3x - 1)\,,&1/3\leq x < 2/3,\\
      \snp{n} + \snz{n} + \snm{n} y_{n-1}(3x-2)\,,&2/3 \leq x < 1,
    \end{array}
  \right.
  \nonumber
\end{eqnarray}
and
\begin{eqnarray}
  \lefteqn{F_n(x) =}
  \label{Hx}\\
  && \left\{
    \begin{array}{l@{\quad}c}
      \frac{y_n(x)}{2 \snp{n}} + \snm{n+1} F_{n+1}(3 x)\,,
      &0\leq x < 1/3,\\   
      \frac{1}{2} + \snz{n} F_n(3 x - 1)\,,
      &{1}/{3}\leq y <{2}/{3},\\ 
      \frac{1 - y_n(x)}{2 \snm{n}} +
      \snp{n-1} F_{n - 1}(3 x - 2)\,,  
      & {2}/{3} \leq y < 1.
    \end{array}
  \right.
  \nonumber
\end{eqnarray}
The boundary conditions are taken so that $y_n(x) = y_1(x)$, $n < 1$, and
$y_n(x) = y_{N}(x)$, $n > N$. As above, $F_n(x) = 0$, $n < 1$ or
$n> N$.
  
Starting from the end points $y(n,0) = 0$, $y(n,1) = 1$ and
$F_n(0) = F_n(1) = 0$, $1\leq  n \leq N$, we
successively compute $y_n(x_k)$ and $F_n(x_k)$, $1\leq n \leq N$ at points
$x_k = \sum_{j=1}^{k} 3^{-j}\omega_j$, where, for every $k\geq 1$, there
are $3^k$ different  sequences $\{\omega_1, \dots, \omega_k\}$,
$\omega_j\in\{0,1,2\}$, $1\leq j \leq k$. 

\begin{figure*}[htp]
  \centering
  \includegraphics[angle=0,width=.48\textwidth]{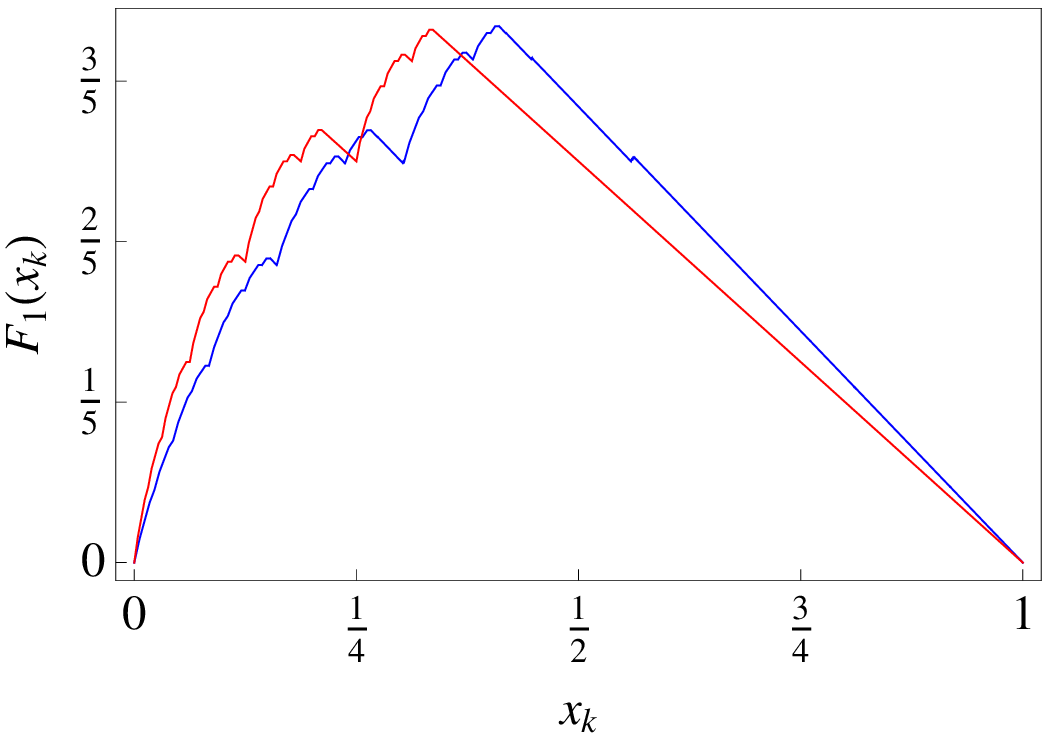}
  \hfill
  \includegraphics[angle=0,width=.48\textwidth]{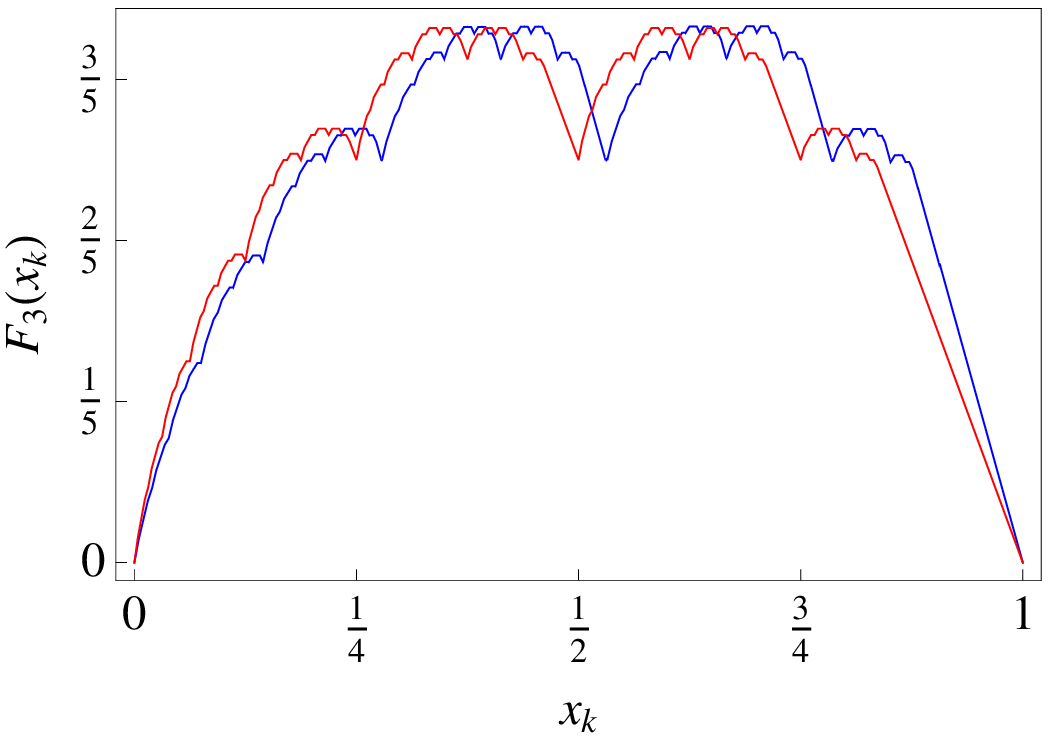}
  \includegraphics[angle=0,width=.48\textwidth]{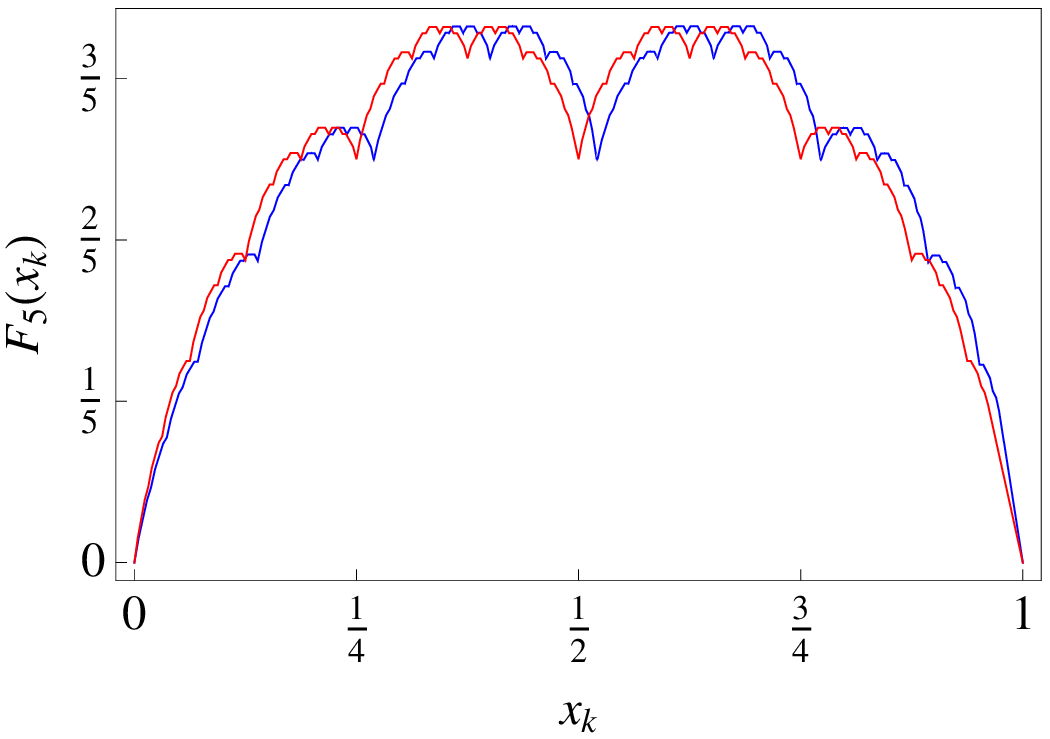}
  \hfill
  \includegraphics[angle=0,width=.48\textwidth]{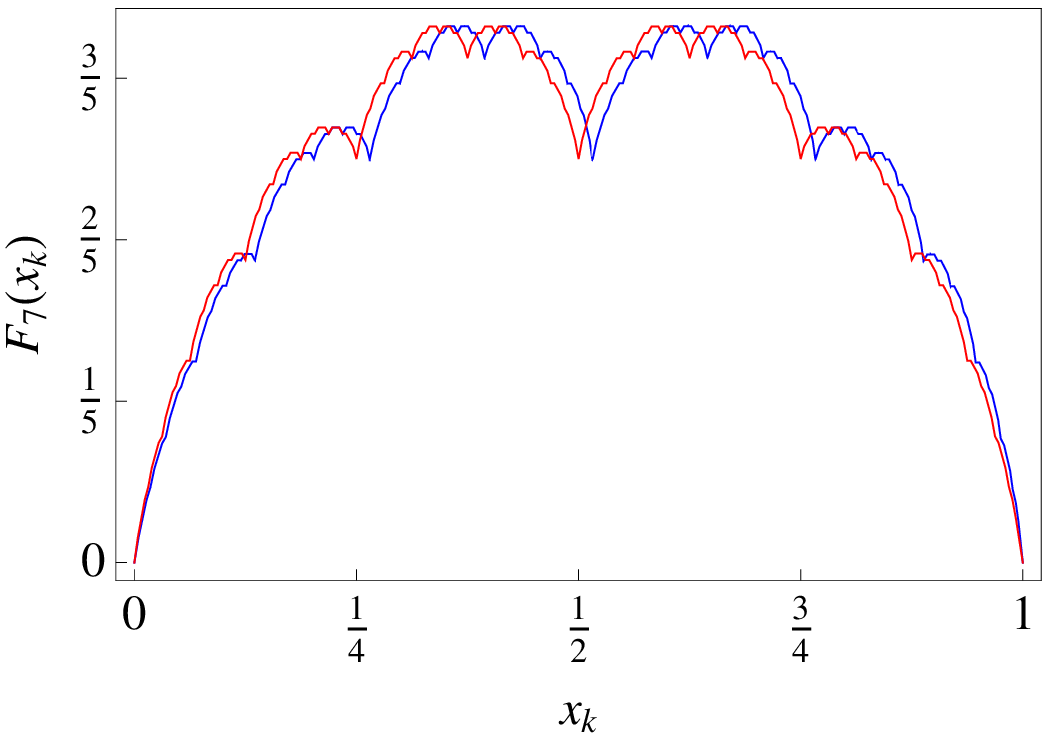}
  \includegraphics[angle=0,width=.48\textwidth]{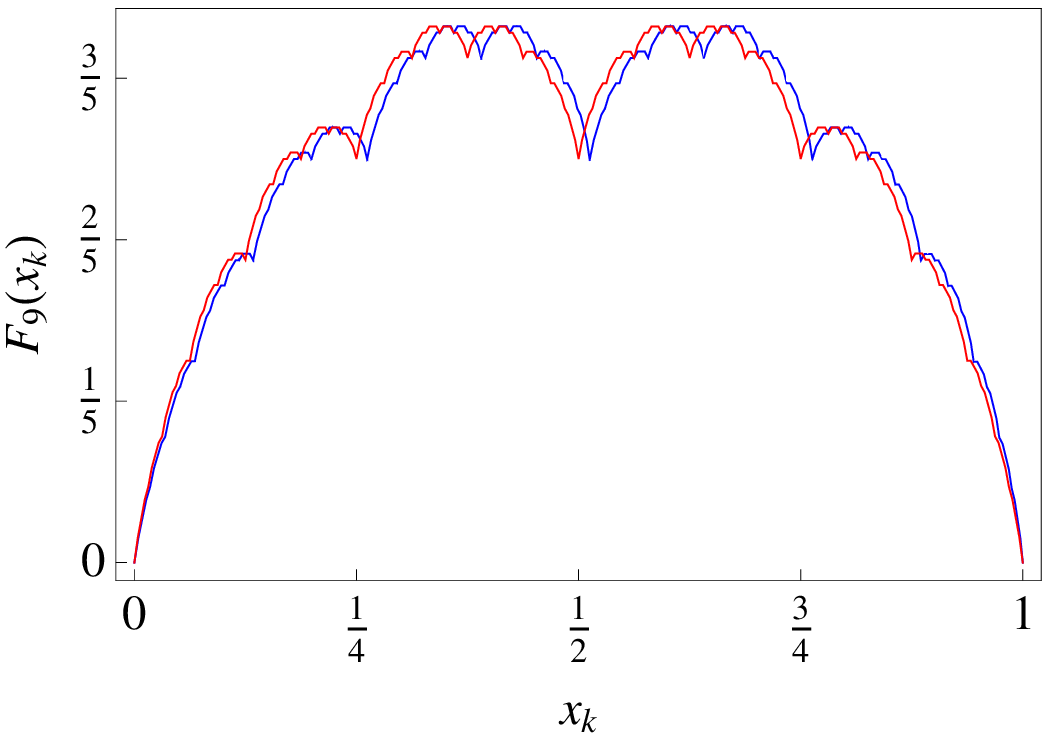}
  \hfill
  \includegraphics[angle=0,width=.48\textwidth]{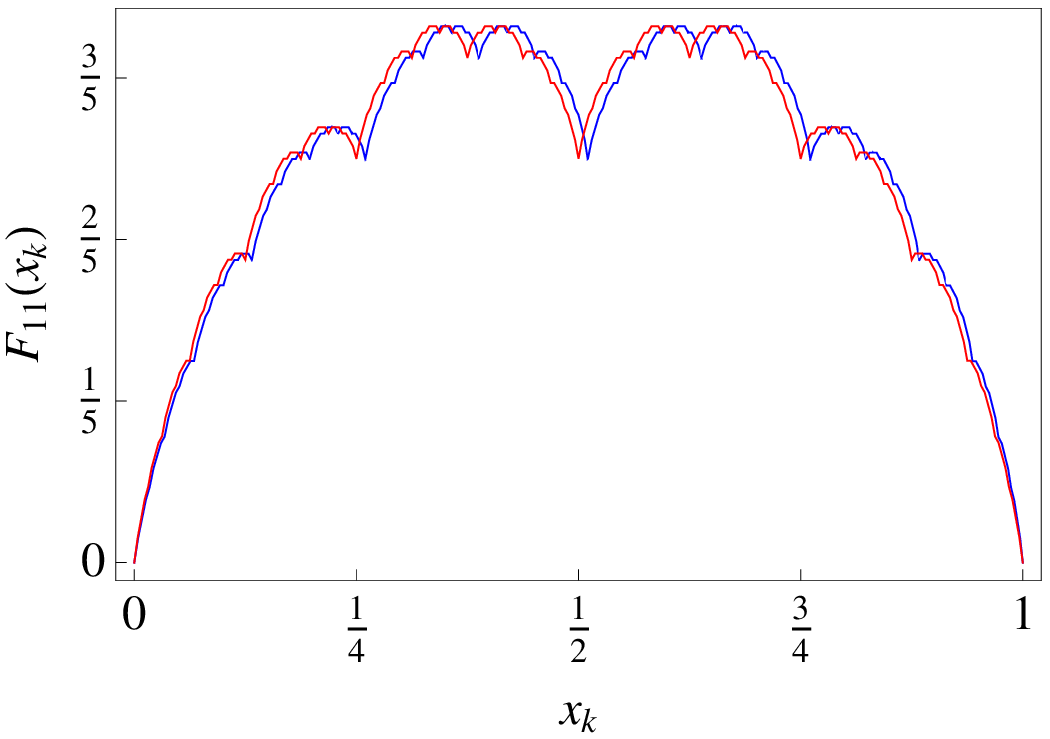}
  \includegraphics[angle=0,width=.48\textwidth]{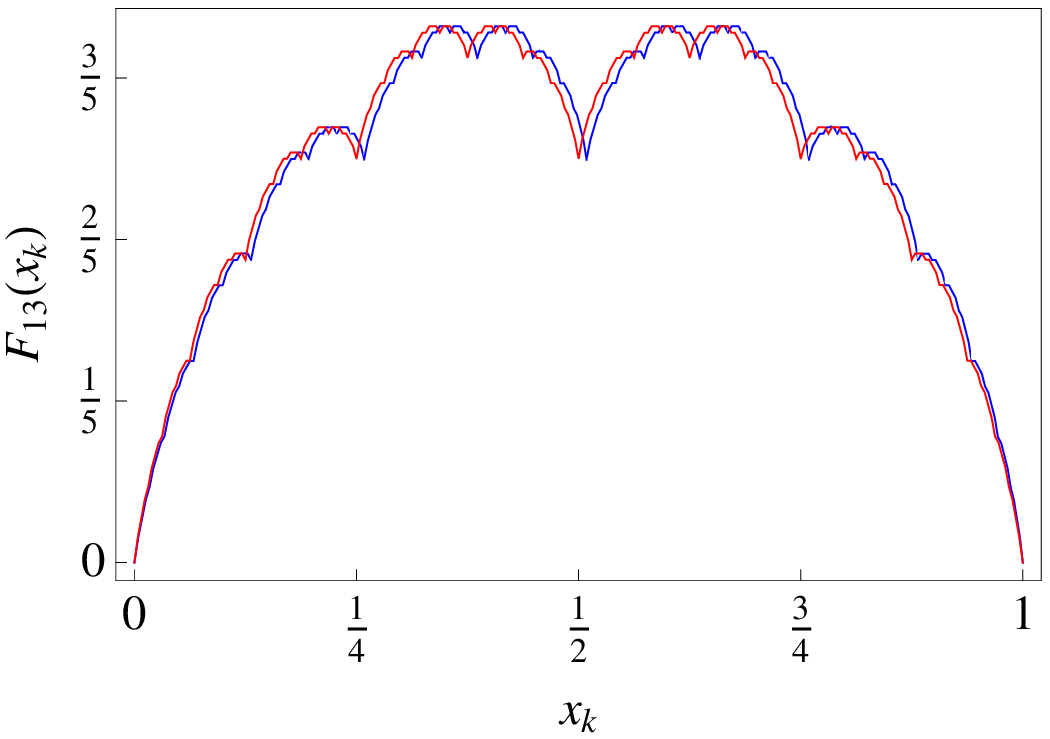}
  \hfill
  \includegraphics[angle=0,width=.48\textwidth]{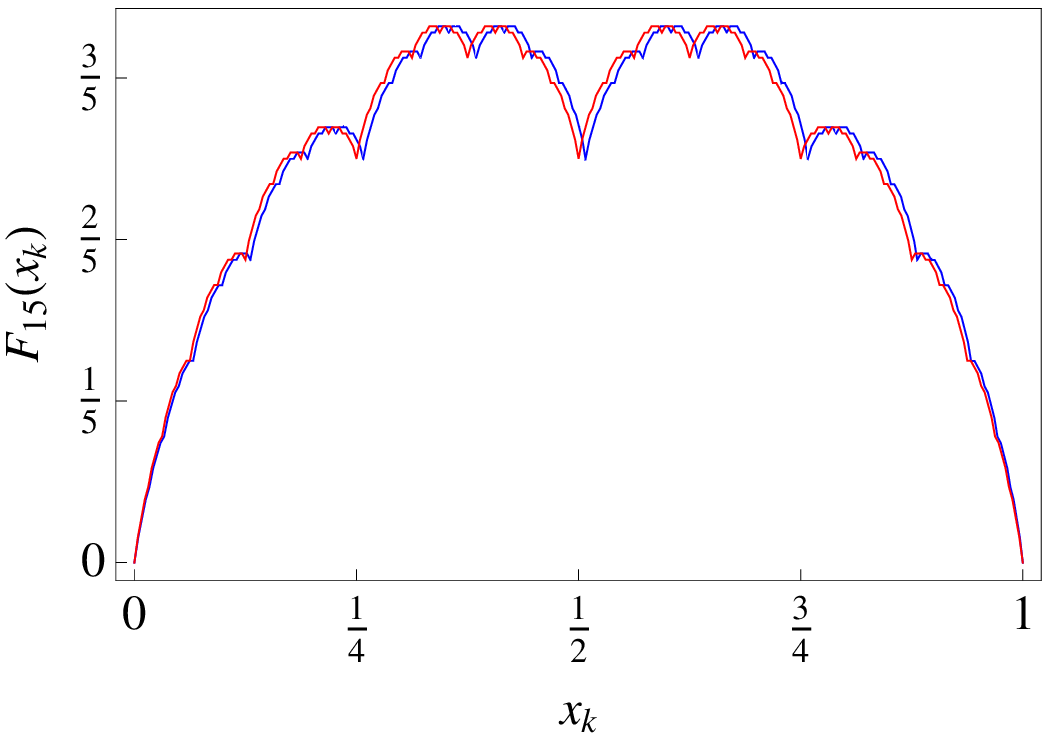}
  \caption{(Color online) Comparison between the graphs of $F_n(x_k)$
    vs. $y_n(x_k)$ (Blue) and the corresponding incomplete Takagi functions
    $T_n(x_k)$ (Red)\,. Each curve is computed at $3^{10}+1$ different points
    $x$, uniformly spread between   $0$ and $1$. Only $2^{10}+1$
    correspond to different points in the graphs of $T_n$.}
  \label{fig.Hy}
\end{figure*}

The graphs of $F_n(x_k)$ vs. $y_n(x_k)$ are displayed in
Fig.~\ref{fig.Hy} for a chain of $N=100$ sites and $k=8$, and compared to
the corresponding graphs of the incomplete Takagi functions \cite{TG95},
which can be obtained from Eq.~(\ref{Hx}) by setting $\snp{n} = \snm{n}
\equiv 1/2$ and $\snz{n} = 0$,
\begin{equation}
y(x) = \left\{
\begin{array}{l@{\quad}c}
\frac{1}{2} y(3 x)\,, &0\leq x < {1}/{3},\\
\frac{1}{2}, &{1}/{3}\leq x < {2}/{3},\\
\frac{1}{2} + \frac{1}{2} y(3x-2)\,,&{2}/{3} \leq x < 1,
\end{array}
\right.
\label{yxT}
\end{equation}
and
\begin{eqnarray}
  \lefteqn{T_n(x) =}
  \label{Tx}\\
  &&\left\{
    \begin{array}{l@{\quad}c}
      y(x) + \frac{1}{2} T_{n+1}(3 x)\,,
      &0\leq x < {1}/{3},\\   
      \frac{1}{2} ,
      &{1}/{3}\leq y <{2}/{3},\\ 
      1 - y(x) +
      \frac{1}{2} T_{n - 1}(3 x - 2)\,,  
      & {2}/{3} \leq y < 1.
    \end{array}
  \right.
  \nonumber
\end{eqnarray}
In passing, we note that, on the one hand, Eq.~(\ref{yxT}) is a functional
equation whose solution is the Cantor function. On the other hand, the
tri-adic representation of the incomplete Takagi functions,
Eq.~(\ref{Tx}), is many-to-one. Their graphs, $T_n(x)$ vs. $y(x)$, are
nevertheless identical to those obtained using the usual representation of
the incomplete Takagi functions.

\subsubsection{Symbolic dynamics}

By substituting the tri-adic expansion of $x$ in Eqs.~(\ref{yx})-(\ref{Hx}),
$x(\{\omega_0,\dots,\omega_k\}) = \sum_{j=0}^{k} \omega_j 3^{-(j+1)}$,
$\omega_j \in \{0,1,2\}$, we obtain the following symbolic representations
of points $y$ in cell $n$,
\begin{eqnarray}
  \lefteqn{y_n(\{\omega_0,\dots,\omega_k\}) =}
  \\
  &&
  \left\{
    \begin{array}{l@{\quad}c}
      \snp{n} y_{n+1}(\{\omega_1,\dots,\omega_k\})\,,&\omega_0 = 0,\\
      \snp{n} + \snz{n} y_n(\{\omega_1,\dots,\omega_k\})\,,&\omega_0 = 1,\\
      \snp{n} + \snz{n} + \snm{n} y_{n-1}(\{\omega_1,\dots,\omega_k\})\,,
      &\omega_0
      = 2. 
    \end{array}
  \right.
  \nonumber
\end{eqnarray}
Starting from 
\begin{equation}
  y_n(\{\omega_0\}) = 
  \left\{
    \begin{array}{l@{\quad}c}
      0,&\omega_0 = 0,\\
      \snp{n},&\omega_0 = 1,\\
      \snp{n}+\snz{n},&\omega_0 = 2,
    \end{array}
  \right.
\end{equation}
we can write
\begin{eqnarray}
  \lefteqn{y_n(\{\omega_0,\dots,\omega_k\}) =}\nonumber\\
  && y_n(\{\omega_0\}) + 
  s_n^{1-\omega_0} y_{n + 1-\omega_0}(\{\omega_1,\dots,\omega_k\})\,,
  \nonumber\\
  &=&y_n(\{\omega_0\}) + s_n^{1-\omega_0} y_{n + 1 - \omega_0}(\{\omega_1\})
  \nonumber\\
  &&+
  s_n^{1-\omega_0} s_{n + 1 - \omega_0}^{1 - \omega_1} 
  y_{n + 2 - \omega_0 -\omega_1}(\{\omega_2, \dots, \omega_k\})\,,
  \nonumber\\
  &\vdots&\\
  &=& \sum_{i=0}^k \left[\prod_{j=0}^{i-1} s_{n + j - \omega_0 - \dots -
      \omega_{j-1}}^{1 - \omega_j} \right] y_{n + i - \omega_0 - \dots -
    \omega_{i-1}}(\{\omega_i\})\,.
  \nonumber
\end{eqnarray}
Substituting this symbolic dynamics into the expression of $F_n$,
Eq. (\ref{ssbmh}), we write
\begin{widetext}
\begin{equation}
  F_n(\{\omega_0,\dots,\omega_k\})
  =
  \left\{
    \begin{array}{l@{\quad}c}
     \frac{1}{2} y_{n + 1}(\{\omega_1, \dots, \omega_k\}) +
      \snm{n+1} F_{n+1}(\{\omega_1,\dots,\omega_k\})\,,& 
      \omega_0 = 0,\\   
      1/2 + \snz{n} F_n(\{\omega_1,\dots,\omega_k\})\,,
      & \omega_0 = 1,\\ 
      \frac{1}{2}(1 - y_{n - 1}(\{\omega_1,\dots,\omega_k\})) +
      \snp{n-1} F_{n - 1}(\{\omega_1,\dots,\omega_k\})\,,  
      & \omega_0 = 2.
    \end{array}
  \right.
  \label{ssbmhsymb}
\end{equation}
\end{widetext}

Let $\Delta y_n(\omega_0,\dots,\omega_k)$ denote the height of a
horizontal cylinder set of the unit square, coded by the sequence
$\{\omega_0,\dots,\omega_k\}$. We have
\begin{eqnarray}
  \lefteqn{\Delta y_n(\omega_0,\dots,\omega_k) \equiv}
  \\
  &&  
  y_n(\{\omega_0,\dots,\omega_k+1\}) -  y_n(\{\omega_0,\dots,\omega_k\})
  \,,
  \nonumber
\end{eqnarray}
where the notation $y_n(\{\omega_0,\dots,\omega_k+1\})$ is literal whenever
$\omega_k \neq 2$. Otherwise  $y_n(\{\omega_0,\dots,\omega_{k-1}, 2+1\})
\equiv y_n(\{\omega_0,\dots,\omega_{k-1}+1,0\})$ and we set
$y_n(\{2,\dots,2,2+1\}) \equiv 1$.
We have the following identities
\begin{equation}
  \left\{
    \begin{array}{lcl}
      \Delta y_n(0, \omega_1,\dots,\omega_k) &=&
      \snp{n}\Delta y_{n + 1}(\omega_1,\dots,\omega_k)\,,
      \nonumber\\
      \Delta y_n(1, \omega_1,\dots,\omega_k) &=&
      \snz{n}\Delta y_n(\omega_1,\dots,\omega_k)\,,
      \\
      \Delta y_n(2, \omega_1,\dots,\omega_k) &=&
      \snm{n}\Delta y_{n - 1}(\omega_1,\dots,\omega_k)\,.
    \end{array}
  \right.
  \label{deltayrecursion}
\end{equation}
Therefore
\begin{equation}
  \Delta y_n(\omega_0,\dots,\omega_k) = \prod_{i=0}^k s_{n + i -
    \omega_0 - \dots - \omega_{i-1}}^{1-\omega_i},
  \label{deltay}
\end{equation}
which is nothing but the probability associated to the trajectory starting
at position $n$ and coded by the sequence $\{\omega_0, \dots, \omega_k\}$.

Likewise, the measure of the cylinder set $\Delta y_n(\omega_0, \dots,
\omega_k)$ is 
\begin{eqnarray}
  \lefteqn{\Delta \mu_n (\omega_0, \dots, \omega_k) \equiv
    \mu_n\big(y_n(\{\omega_0, \dots, \omega_{k - 1}+1\})\big)}
  \nonumber\\
  &&- 
  \mu_n\big(y_n(\{\omega_0, \dots, \omega_{k - 1}\})\big)
  \,,\nonumber\\
  &=& \mu_n \Delta y_n(\omega_0, \dots, \omega_k) 
  + \alpha \Delta F_n(\omega_0, \dots, \omega_k)\,, 
\end{eqnarray}
and we have the following
set of identities for $\Delta F_n$:  
\begin{widetext}
\begin{equation}
  \left\{
    \begin{array}{lcl}
      \Delta F_n(0, \omega_1,\dots,\omega_k) &=&
      \frac{1}{2}\Delta y_{n + 1}(\omega_1,\dots,\omega_k)
      + \snm{n+1} \Delta F{n + 1}(\omega_1,\dots,\omega_k)\,,
      \\
      \Delta F_{n}(1, \omega_1,\dots,\omega_k) &=&
      \snz{n}\Delta F_n(\omega_1,\dots,\omega_k)\,,\\
      \Delta F_n(2, \omega_1,\dots,\omega_k) &=&
      -\frac{1}{2}\Delta y_{n - 1}(\omega_1,\dots,\omega_k)
      + \snp{n-1} \Delta F_{n - 1}(\omega_1,\dots,\omega_k)\,.
    \end{array}
  \right.
  \label{deltaHrecursion}
\end{equation}
That is,
\begin{equation}
  \Delta F_n(\omega_0,\dots,\omega_k) =
  \frac{1}{2}(1-\omega_0) \Delta y_{n + 1-\omega_0}
  (\omega_1,\dots,\omega_k) + s_{n+1-\omega_0}^{\omega_0-1} \Delta
  F_{n + 1 - \omega_0}(\omega_1,\dots,\omega_k)\,.
  \label{deltaH}
\end{equation}
\end{widetext}
Notice that it is possible to solve this system recursively, starting from
 \begin{equation}
   \Delta F_n(\omega_0) =
   \left\{
     \begin{array}{l@{\quad}c}
       1/2\,,&\omega_0 = 0\,,\\
       0\,,&\omega_0 = 1\,,\\
       - 1/2\,,&\omega_0 = 2\,.
     \end{array}
   \right.
 \end{equation}
We thus have a complete characterization of the non-equilibrium stationary
state of $B$, Eq.~(\ref{fmbmap}), associated to flux boundary conditions.
%\end{widetext}

\subsection{\label{subsec.ep}Entropy and Entropy Production}

We proceed along the lines of \cite{Gas97a, GD99} to obtain expressions of the
entropies and entropy production rates associated to coarse grained sets
such as defined in Eq.~(\ref{deltay}). As described in \cite{BGGi}, the
idea is that, owing to the singularity of the invariant density,  the
entropy should be defined with respect to a grid of phase space, or
partition, $\mathbb{G} = \{d\Gamma_j\}$, into small volume elements
$d\Gamma_j$, and a time-dependent state $\mu_n(d\Gamma_j, t)$. The entropy
associated to cell $\mathbb{C}_n$, coarse grained with respect that grid, is
defined according to 
\begin{equation}
  S_{\mathbb{G}}^t(\mathbb{C}_n) = - \sum_j
  \mu_n(d\Gamma_j, t) \left[\log\frac{\mu_n(d\Gamma_j, t)}{d\Gamma_j} -
    1\right]. 
  \label{Gibbsentcc}
\end{equation}
This entropy changes in a time interval $\tau$ according to
\begin{eqnarray}
  \Delta^\tau S^t(\mathbb{C}_n) 
  &=& S_{\mathbb{G}}^t(\mathbb{C}_n) - S_{\mathbb{G}}^{t-\tau}(\mathbb{C}_n)
  \,,\label{deltaStau}\\
  &=& S_{\{d\Gamma_j\}}^t(\mathbb{C}_n) - 
  S_{\{\Phi^\tau d\Gamma_j\}}^{t}(\Phi^\tau \mathbb{C}_n)
  \,,
  \nonumber
\end{eqnarray}
where, in the second line, the collection of partition elements
$\{d\Gamma_j\}$ was mapped to $\{\Phi^\tau d\Gamma_j\}$, which forms a
partition $\Phi^\tau\mathbb{G}$ whose elements are typically stretched
along the unstable foliations and folded along the stable foliations. 

Following \cite{DGG02}, and in a way analogous to the phenomenological
approach to entropy production \cite{deGMaz}, the rate of entropy change
can be further decomposed into entropy flux and production terms according
to
\begin{equation}
  \Delta^\tau S_{\mathbb{G}}^t(\mathbb{C}_n) = 
  \Delta_\mathrm{e}^\tau S_{\mathbb{G}}^t(\mathbb{C}_n) 
  + \Delta_\mathrm{i}^\tau S_{\mathbb{G}}^t(\mathbb{C}_n)\,,
  \label{entdecomp}
\end{equation}
where the entropy flux is defined as the difference between the entropy
that enters cell $\mathbb{C}_n$ and the entropy that exits that cell,
\begin{equation}
  \Delta_\mathrm{e}^\tau S_{\mathbb{G}}^t(\mathbb{C}_n) = 
  S_{\{\Phi^\tau d\Gamma_j\}}^t(\mathbb{C}_n) 
  - S_{\{\Phi^\tau d\Gamma_j\}}^t(\Phi^\tau \mathbb{C}_n) \,.
  \label{entchange}
\end{equation}
Collecting Eqs.~(\ref{deltaStau})-(\ref{entchange}), the entropy production
rate at $\mathbb{C}_n$ measured with respect to the partition $\mathbb{G}$,
is identified as
\begin{equation}
  \Delta_\mathrm{i}^\tau S_{\mathbb{G}}^t(\mathbb{C}_n) = 
  S_{\{d\Gamma_j\}}^t(\mathbb{C}_n) 
  - S_{\{\Phi^\tau d\Gamma_j\}}^t(\mathbb{C}_n) \,.
  \label{entprod}
\end{equation}
This formula is equally valid in the non-equilibrium stationary state.

Given a phase-space partition into the $3^k$ cylinder sets coded by the
sequences $\uo{k}\equiv \{\omega_0,\dots,\omega_{k-1}\}$,
$\omega_i\in\{0,1,2\}$, as described in Sec. \ref{subsec.ste}, 
the $k$-entropy of the stationary state Eq. (\ref{sssubs}) relative to the
volume measure of cell $n$ is defined by  
\begin{equation}
S_k(\mathbb{C}_n) = - \sum_{\uo{k}}
\Delta \mu_n(\uo{k}) \left[\log
\frac {\Delta \mu_n(\uo{k})} 
{\Delta y_n(\uo{k})}-1\right]\,.
\label{kent}
\end{equation}

By summing over the first digit, it follows immediately from
Eqs.~(\ref{pfbm}) and (\ref{deltay}) that the $k$-entropy verifies a
recursion relation,
\begin{eqnarray}
  S_k(\mathbb{C}_n) &=& - \snm{n+1} \rho_{n+1} \log \frac{\snm{n+1}}{\snp{n}}
  - \snp{n-1} \rho_{n-1} \log \frac{\snp{n-1}}{\snm{n}}\nonumber\\
  &&+ \snm{n+1} S_{k-1}(\mathbb{C}_{n+1}) 
  + \snz{n} S_{k-1}(\mathbb{C}_n) \nonumber\\
  &&+ \snp{n-1} S_{k-1}(\mathbb{C}_{n-1})\,,
\label{kentrecur}
\end{eqnarray}
with the $0$-entropy given by
\begin{equation}
S_0(\mathbb{C}_n) = -\mu_n \log \mu_n\,,
\label{0ent}
\end{equation}
and boundary conditions
\begin{equation}
\left\{
\begin{array}{lcl}
S_k(\mathbb{C}_0) &=& -\rho_- \log \rho_-\,,\\
S_k(\mathbb{C}_{N+1}) &=& -\rho_+ \log \rho_+\,.
\end{array}
\right.
\label{BCkent}
\end{equation}
The $k$-entropy can be computed based on the above recursion relation.
However, in order to obtain the dependence of the entropy on the resolution
parameter $k$, it is more useful to consider the expansion of
Eq.~(\ref{kent}) in powers of $\rho_n$. Let us denote by $\uo{k}$ the
sequence $\{\omega_0, 
\dots, \omega_{k-1}\}$ 
\begin{widetext}
\begin{eqnarray}
  S_k(\mathbb{C}_n) &=& 
  - \mu_n \sum_{\uo{k}}\Delta y_n(\uo{k})
  \left[ 1 + \frac{\alpha}{\mu_n}\frac {\Delta F_n(\uo{k})}
    {\Delta y_n(\uo{k})} \right] \left\{\log \mu_n
  \left[ 1 + \frac{\alpha}{\mu_n}\frac {\Delta F_n(\uo{k})}
    {l \Delta y_n(\uo{k})} \right]-1\right\}\,, \nonumber\\
  &=& - \mu_n (\log \mu_n - 1)
  - \alpha (\log \mu_n - 1) \sum_{\uo{k}} \Delta F_n(\uo{k})
  - \frac{\alpha^2} {2\mu_n} \sum_{\uo{k}} \frac 
  {[\Delta F_n (\uo{k})]^2}{\Delta y_n(\uo{k})}
  + \mathcal{O}(\alpha)^{3}\,.
  \label{kentexpand}
\end{eqnarray}
The second term on the RHS of this equation vanishes, since 
\begin{equation}
\sum_{\uo{k}}\Delta F_n(\uo{k}) = 0.
\label{sumDeltaH}
\end{equation}
As of the third term on the RHS of Eq.~(\ref{kentexpand}), proportional to
$\alpha^2$, we have, using Eqs.~(\ref{deltayrecursion}) and
(\ref{deltaHrecursion}), 
\begin{eqnarray}
  \Delta^2_n(k) &\equiv&
  \sum_{\uo{k}}\frac {[\Delta F_n(\uo{k})]^2}{\Delta y_n(\uo{k})}\,,
  \nonumber\\
  &=& \frac{1}{4 \snp{n}} + \frac{1}{4 \snm{n}}
  + \frac{(\snm{n + 1})^2}{\snp{n}} \Delta^2_{n + 1}(k-1)
  + \snz{n} \Delta^2_{n}(k-1)
  + \frac{(\snp{n - 1})^2}{\snm{n}} \Delta^2_{n - 1}(k-1)
  \nonumber\\
  &=& \frac{1}{4 \snp{n}} + \frac{1}{4 \snm{n}}
  + \sum_{\eta=0}^2 \frac{(s^{\eta-1}_{n + 1 - \eta})^2}{s^{1-\eta}_{n}} 
  \Delta^2_{n + 1 - \eta}(k-1)
  ,\label{Delta2recurs}\\
  &=&
  \sum_{i=0}^{k - 1} \sum_{\eta_1,\dots,\eta_{i - 1}}
  \left[\prod_{j = 1}^{i} \frac 
    {(s^{\eta_j - 1}_{n + j - \eta_1 - \dots - \eta_j})^2}
    {s^{1-\eta_j}_{n + j - 1 - \eta_1 - \dots - \eta_{j-1}}}\right]
  \left(\frac{1}{4 \snp{n + i - \eta_1 - \dots - \eta_i}} + 
    \frac{1}{4 \snm{n + i - \eta_1 - \dots - \eta_i}} \right)\,.
  \label{Delta2recursex}
\end{eqnarray}
\end{widetext}
Substituting the expressions of the probability transitions from 
Eqs.~(\ref{rsnexpand})-(\ref{rsnzexpand}), Eq.~(\ref{Delta2recursex}) is
found to be  
\begin{eqnarray}
  \Delta^2_n(k) &=& 
  \sum_{\uo{k}}\frac {[\Delta F_n(\uo{k})]^2}{\Delta y_n(\uo{k})}\,,
  \nonumber\\
  &=& 
  k + \frac{12 k^2 - 9 k}{32 (n+n_0)^2} + \mathcal{O}(n+n_0)^{-4}\,.
  \label{sumdeltaHsq}
\end{eqnarray}
The first term on the RHS of this expression, which is the only term that
survives in the continuum limit where $n+n_0\gg1$, is responsible for the
linear decay of the $k$-entropy,
\begin{equation}
  S_k(\mathbb{C}_n) \simeq - \mu_n (\log \mu_n - 1)
  - \frac{\alpha^2} {2\mu_n}k\,.
  \label{kentfinal}
\end{equation}

Indeed if follows from Eq. (\ref{entprod}) that the $k$-entropy production
rate is here
\begin{eqnarray}
  \Delta_\mathrm{i}^\tau S_k(\mathbb{C}_n) &=& \frac{1}{\tau} \left[
    S_k(\mathbb{C}_n)  - 
    S_{k+1}(\mathbb{C}_n) \right],
  \label{fmbmapkep}\\
  &=& \frac{\alpha^2}{2\tau \mu_n}\,.
  \nonumber
\end{eqnarray}
Using Eqs.~(\ref{fDiff}), (\ref{ssFrPeq}) and (\ref{alpha}), it is readily
checked that this expression yields the 
phenomenological entropy production rate, Eq.~(\ref{gbFPep}),
$\Delta_\mathrm{i}^\tau S_k(\mathbb{C}_n) \stackrel{\tau,l\to0}{\to}
d_\mathrm{i} S(X =  nl)/dt$. 

\section{\label{sec.con}Conclusions}

In this paper, we have considered the influence of an external
field on a class of time-reversible deterministic volume-preserving models
of diffusive systems known as Galton boards or, equivalently, forced
periodic two-dimensional Lorentz gases.  

Though the particles are accelerated as they move along the direction of
the external field, in the absence of a dissipative mechanism, the motion is
recurrent, which is to say that tracer particles keep coming back to the
region of near zero velocity. In other words, particles do not drift in the
direction of the external field. Rather, forced periodic Lorentz gases
remain purely diffusive in two dimensions, albeit with a velocity-dependent
diffusion coefficient. Consequently, the scaling laws relating time and
displacement are different from that of a homogeneously diffusive
system. The macroscopic description through a Fokker Planck equation is
however unchanged since the mobility coefficient vanishes identically in
dimension 2. 

It will be interesting to investigate the behavior of three-dimensional
periodic Lorentz gases in a uniform external field. As our analysis showed,
the mobility does not vanish in dimension three, so that the Fokker-Planck
equation retains a drift term. Being inversely proportional to the tracers'
velocity amplitudes, this drift decreases with increasing kinetic energy. A
cross-over is thus expected between biased and diffusive motions.

As far as their statistical properties are concerned, Galton boards are
essentially identical to the field-free periodic two-dimensional Lorentz
gases. A closed system with reflecting boundaries relaxes to an equilibrium
state with a uniform invariant measure. This is to say that tracers spend
equal amounts of time in all parts of the system. Open systems with
absorbing boundaries yield non-equilibrium states. Given constant rates of
tracer injection at the borders, the system reaches a non-equilibrium
stationary state which is characterized by a fractal invariant measure.

The fractality of the invariant measure associated to the non-equilibrium
state of such a system was established analytically for a multi-baker map
describing the motion of random walkers accelerated by a uniform external
field. The computation of the coarse grained entropies associated to
arbitrarily refined partitions yields expressions which depart from their
local equilibrium expressions by a term which decreases linearly with the
logarithm of the number of elements in the partition. This term is
responsible for the positiveness of the entropy production rate, with a value
consistent with the phenomenological expression of thermodynamics.

%-------------------------------------------------------------------------------
\begin{acknowledgments}
This research is financially supported by the Belgian Federal 
Government  (IAP project ``NOSY") and the ``Communaut\'e fran\c caise de
Belgique'' (contract ``Actions de Recherche Concert\'ees''
No. 04/09-312) as well as by the Chilean Fondecyt under 
International Cooperation Project 7070289. 
TG is financially supported by the Fonds de la Recherche
Scientifique F.R.S.-FNRS. FB acknowledges financial support from the 
Fondecyt Project 1060820 and FONDAP 11980002 and Anillo ACT 15. 
\end{acknowledgments}
%-------------------------------------------------------------------------------

\end{document}